\PassOptionsToPackage{hyphens}{url}
\documentclass[manuscript,screen]{acmart}
\AtBeginDocument{%
  }

\setcopyright{acmlicensed}
\copyrightyear{2025}
\acmYear{2025}
\acmDOI{XXXXXXX.XXXXXXX}

\usepackage{natbib}
\usepackage{soul}
\usepackage{enumitem}
\usepackage{balance}
\usepackage{framed}
\usepackage{acronym}
\usepackage{amsmath}
\usepackage{algorithmic}
\usepackage{graphicx}
\usepackage{textcomp}
\usepackage{xcolor}
\usepackage{xspace}
\usepackage[many]{tcolorbox}
\usepackage{totcount}
\usepackage{multirow}
\usepackage{longtable}

\graphicspath{ {./figs/} }

\newcommand{\comp}{compiler\xspace}
\newcommand{\comps}{compilers\xspace}
\newcommand{\compilation}{compilation\xspace}
\newcommand{\compilations}{compilations\xspace}
\newcommand{\compiling}{compiling\xspace}
\newcommand{\compile}{compile\xspace}
\newcommand{\compiled}{compiled\xspace}

\newcommand{\Compilation}{Compilation\xspace}
\newcommand{\Compiling}{Compiling\xspace}

\newcommand{\autocog}{\textit{Compiler.next}\xspace}
\newcommand{\fmware}{FMware\xspace}

\newcommand{\agentware}{Agentware\xspace}
\newcommand{\codeware}{Codeware\xspace}
\newcommand{\promptware}{Promptware\xspace}
\newcommand{\neuralware}{Neuralware\xspace}

\newcommand{\challenge}{call for action\xspace}
\newcommand{\challenges}{calls for action\xspace}
\newcommand{\Challenge}{Call for Action\xspace}

\newcommand{\langchain}{{LangChain}\xspace}
\newcommand{\autogen}{{AutoGen}\xspace}
\newcommand{\dspy}{{DSPy}\xspace}
\newcommand{\sammo}{{SAMMO}\xspace}
\newcommand{\ape}{{APE}\xspace}
\newcommand{\promptbreeder}{{Promptbreeder}\xspace}
\newcommand{\textgrad}{{TextGrad}\xspace}
\newcommand{\evoprompt}{{EvoPrompt}\xspace}
\newcommand{\protegi}{{ProTeGi}\xspace}
\newcommand{\pytorch}{{PyTorch}\xspace}

\newcommand{\constructs}{constructs\xspace}

\newcommand{\nsgii}{NSGA-II\xspace}

\newboolean{showcomments}
\setboolean{showcomments}{true}

\ifthenelse{\boolean{showcomments}}
{\newcommand{\nbc}[3]{
 {\colorbox{#3}{\bfseries\sffamily\scriptsize\textcolor{white}{#1}}}
 {\textcolor{#3}{\sf\small$\blacktriangleright$\textit{#2}$\blacktriangleleft$}}
 }
}
{\newcommand{\nbc}[3]{}
 }

\newcommand\prompt[1]{\texttt{#1}}
\newcommand\class[1]{\texttt{#1}}

\newcounter{count}
\newcommand\cfa[1]{%
    \refstepcounter{count}%
    \leftbar
    \noindent \textbf{\Challenge \thecount:} #1
    \endleftbar
}

\regtotcounter{count}
\newcommand{\ncfas}{\total{count}\xspace}

\newtcolorbox{challengebox}[2][]{
top=0.15in,left=4pt,right=4pt,bottom=4pt,
fonttitle=\bfseries,
colbacktitle=gray,
colback=white,
colframe=gray!40!black,
enhanced,
breakable,
attach boxed title to top left={xshift=1.5em,yshift=-\tcboxedtitleheight/2},
boxed title style={size=small},
drop shadow={black!50!white},
title=#2,#1}

\acrodef{fm}[FM]{Foundation Model}
\acrodefplural{fm}[FMs]{Foundation Models}
\acrodef{dl}[DL]{deep learning}
\acrodef{se}[SE]{software engineering}
\acrodef{ai}[AI]{artificial intelligence}
\acrodef{rag}[RAG]{retrieval-augmented generation}
\acrodef{ca}[CA]{cognitive architecture}
\acrodefplural{ca}[CAs]{cognitive architectures}
\acrodef{ir}[IR]{intermediate representation}
\acrodefplural{ir}[IRs]{intermediate representations}
\acrodef{randd}[R\&D]{research and development}
\acrodef{db}[DB]{database}
\acrodef{ml}[ML]{machine learning}
\acrodef{cot}[CoT]{chain-of-thought}
\acrodef{api}[API]{Application Programming Interface}
\acrodefplural{api}[APIs]{Application Programming Interfaces}
\acrodef{dag}[DAG]{direct acyclic graph}
\acrodefplural{dag}[DAGs]{direct acyclic graphs}
\acrodef{pl}[PL]{programming language}
\acrodefplural{pl}[PLs]{programming languages}
\acrodef{ide}[IDE]{Integrated Development Environment}
\acrodef{ga}[GA]{genetic algorithm}
\acrodefplural{ga}[GAs]{genetic algorithms}
\acrodef{se30}[SE 3.0]{Software Engineering 3.0}

\begin{document}

\title{Compiler.next: A Search-Based Compiler to Power the AI-Native Future of Software Engineering}

\author{Filipe R. Cogo}
\email{filipecogo@acm.org}
\orcid{0000-0002-5494-685X}
\affiliation{%
  \institution{Centre for Software Excellence - Huawei Canada}
  \city{Kingston}
  \state{ON}
  \country{Canada}
}

\author{Gustavo A. Oliva}
\email{gustavo.oliva@huawei.com}
\orcid{0000-0002-5419-9284}
\affiliation{%
  \institution{Centre for Software Excellence - Huawei Canada}
  \city{Kingston}
  \state{ON}
  \country{Canada}
}

\author{Ahmed E. Hassan}
\email{ahmed@cs.queensu.ca}
\orcid{0000-0001-7749-5513}
\affiliation{%
  \institution{School of Computing, Queen's University}
  \city{Kingston}
  \state{ON}
  \country{Canada}}


\begin{abstract}
The rapid advancement of AI-assisted software engineering has brought transformative potential to the field of software engineering, but existing tools and paradigms remain limited by cognitive overload, inefficient tool integration, and the narrow capabilities of AI copilots. In response, we propose \autocog, a novel search-based compiler designed to enable the seamless evolution of AI-native software systems as part of the emerging Software Engineering 3.0 era. Unlike traditional static compilers, \autocog takes human-written intents and automatically generates working software by searching for an optimal solution. This process involves dynamic optimization of cognitive architectures and their constituents (e.g., prompts, foundation model configurations, and system parameters) while finding the optimal trade-off between several objectives, such as accuracy, cost, and latency. This paper outlines the architecture of \autocog and positions it as a cornerstone in democratizing software development by lowering the technical barrier for non-experts, enabling scalable, adaptable, and reliable AI-powered software. We present a roadmap to address the core challenges in intent compilation, including developing quality programming constructs, effective search heuristics, reproducibility, and interoperability between compilers. Our vision lays the groundwork for fully automated, search-driven software development, fostering faster innovation and more efficient AI-driven systems.
\end{abstract}



\received{10 January 2025}

\maketitle

\section{Introduction}
\label{sec:introduction}

The rapid rise of \ac{ai}-assisted \ac{se} has significantly transformed the development process, but it has also revealed critical limitations. Developers may face cognitive overload~\cite{vaithilingam2022expectation,jalil2023transformative}, inefficiencies in tool integration~\cite{zhang2023practices,zhou2025exploring}, and the narrow capabilities of \ac{ai} copilots~\cite{jimenez2024swebench,yuan2024evaluating,guo2024exploring}. To overcome these challenges, we have previously proposed a vision of an AI-native future as a representation of a paradigm shift towards \ac{se30}, where human developers and \ac{ai} collaborate seamlessly and their strengths are combined to realize SE tasks~\citep{hassan2024ainativesoftwareengineeringse}.  \ac{se30} redefines software development by moving beyond task-based \ac{ai} assistance towards a paradigm where \ac{ai} systems become intelligent collaborators that are capable of \textbf{autonomously synthesizing human intents into working software} (e.g., an idea into a mobile app, or functional requirements into a system that fulfills a customer's need). We refer to the component responsible for performing this synthesis as \autocog. Autonomous coding agents, despite its current limitations, are now actively initiating, reviewing, and evolving code at scale throughout open-source ecosystems. Recent empirical work provides concrete evidence that AI teammates are already integrated into modern software development workflows~\cite{li2025riseaiteammatessoftware}. Our core contribution in this paper is to provide a systematic framework and research roadmap for this emerging paradigm, particularly the compilation infrastructure needed to transform intents into optimized \fmware, positioning the scattered advances in agent-based development within a coherent vision of AI-native software engineering.

Due to the reasoning and generative capabilities of \acp{fm} like Large Language Models (LLMs), FM-powered systems (\fmware) will grow even more in popularity and become a fundamental type of software in the \ac{se30} era.
However, synthesizing intents into \fmware poses several challenges. \fmware encompasses an entire ecosystem of interdependent components that interact dynamically in real time. This includes retrieval-augmented generation (RAG) to fetch relevant information, continuous data flows powered by a data flywheel approach that provides up-to-date context, and a runtime environment where models are constantly evolving in response to changing inputs, feedback loops, and tasks. As these systems evolve, they require an orchestrated effort to manage these components, ensuring that the system’s performance remains optimal despite the constantly shifting landscape. Therefore, synthesizing \fmware requires more than optimizing prompts with prompt engineering techniques. It requires a whole new, dynamic, integrated approach to generate every component of a complex \ac{ca} while ensuring that those components are aligned and can evolve in harmony.

To synthesize intents into complex \fmware, \autocog employs a search-based approach. By leveraging \acp{fm}, the solution is iteratively developed at hyper speed through code mutations and self-reflection mechanisms that iteratively evaluate how well the resulting software matches the intent. Another key property of \autocog is its ability to find the best trade-off between various competing objectives, such as accuracy (how well it fulfills the intent), latency (e.g., number of API requests sent to one or more \acp{fm}), and cost (e.g., number of input tokens in the prompt). That is, \autocog performs a multi-objective optimization.

Ultimately, \autocog is the realization that \acp{fm} serve as probabilistic CPUs and that prompts act as the ``binaries'' that are executed by these \acp{fm}. By allowing developers to focus on intents (instead of creating and hacking prompts), \autocog powers and accelerates the development of \fmware in the \ac{se30} era by \textbf{searching} for the optional \ac{ca} (e.g., prompts, \ac{fm} selection and configuration, and \ac{rag} components) to fulfill the human's intents while ensuring that multiple criteria are satisfied with the best trade-off possible.

In this paper, we lay out a \textit{research and development roadmap} for \autocog. This roadmap results in \ncfas \textit{call for action} items. We focus specifically on the compilation challenge, i.e. transforming human intents into optimized \fmware through search-based synthesis, as this process directly addresses critical software maintenance challenges. For example, recent research demonstrates that prompts are highly fragile and sensitive to minor variations, with performance varying significantly based on formatting choices~\cite{sclar2024quantifying}, instruction phrasing~\cite{cao2024worstpromptperformancelarge}, and even the order of few-shot examples~\cite{lu2022fantasticallyorderedprompts}. This fragility creates a substantial maintenance burden as FMs evolve and requirements change. By compiling from stable intents, \autocog enables \fmware to adapt to new models and requirements through recompilation, similar to how traditional software can be recompiled for new target architectures. This separation of intent (what to achieve) from implementation (optimized prompts and configurations) is fundamental to sustainable FMware engineering. Therefore, intent compilation represents a foundational capability that enables many downstream SE activities, but it is not the entirety of software engineering in the SE 3.0 era. A complete SE 3.0 ecosystem requires complementary advances in requirements validation, comprehensive testing strategies, FMware deployment and operations, maintenance and evolution processes, and collaborative development workflows. We position \autocog as an enabling technology that must integrate with these broader practices to support complete software development lifecycles. We hope that our set of calls for action will inspire the software engineering community and foster deeper academia-industry discussions and collaborations.

\begin{challengebox}{Calls for action}
    \begin{enumerate}[wide = 0pt, itemsep = 1.5pt, topsep = 1.5pt, label=\bfseries\arabic*.]
         \item Establish quality programming \constructs for representing \fmware programs.
         
         \item Consolidate the forms of \compilation that involves not only prompt templates but all free parameters of an \fmware, including better support for agent-based applications~\cite{hu2024automateddesignagenticsystems}.
         
         \item Identify the set of most effective heuristics to search the space of \fmware parameters.
         
         \item Construct sets of gold labels to evaluate candidate solutions and guide the search procedure during \compilation.
         
         \item Assure the quality of an \fmware application, failing \compilation when quality thresholds are not met.
         
         \item Reduce the cost and improve the efficiency of \comps.
         
         \item Make intent \compilation reproducible.
         
         \item Enable user-defined, multiple concurrent objectives to be optimized during \compilation.
         
         \item Improve the interoperability between \comps.
         
         \item Build community-sharing platforms of compilation traces such that this information can be used as a feedback signal to improve \compilation.
    \end{enumerate}
\end{challengebox}

This paper is structured as follows. Section~\ref{sec:background} discusses background and related work. Section~\ref{sec:compiler-next} describes \autocog, which reflects our vision of a \comp that synthesizes intents into running software via a search-based approach. Section~\ref{sec:calls-for-action} presents a research and development roadmap for \autocog, which includes the \ncfas ``\challenges'' described above. Finally, Section~\ref{sec:conclusions} states our final remarks and concludes the paper.
\section{Background and related work}
\label{sec:background}

Compilers are fundamental to \ac{se} as they transform human-readable code into machine-executable instructions, bridging the gap between developers and computing infrastructure and allowing developers to focus on application logic rather than low-level optimizations. In this section, we describe the role and contributions of traditional (Section~\ref{sec:background:traditional-compilers}), \ac{dl} (Section~\ref{sec:background:deep-learning-compilers}), and prompt (Section~\ref{sec:background:prompt-compilers}) \comps to \ac{se}. 

\subsection{Traditional \comps}
\label{sec:background:traditional-compilers}

Traditional \comps have played a pivotal role in \ac{se} by transforming high-level human-readable code into machine-executable binaries. The primary function of a traditional compiler is to map the abstract syntax of \acp{pl} into the concrete syntax of a target machine, optimizing code at various stages, such as lexical analysis, parsing, and code generation. Over decades, \comps have evolved to include sophisticated optimization techniques like constant folding, loop unrolling, and register allocation, aimed at maximizing performance while minimizing resource utilization~\citep{aho1986compilers}. Despite their enduring importance, traditional \comps operate in a static context: once the code is compiled and optimized, there is no continuous refinement based on runtime feedback.

The rigid architecture of traditional \comps limits their applicability in dynamic, \ac{ai}-powered environments, where systems need to adapt to evolving requirements in real-time. The rise of \ac{ai}-native systems, particularly those powered by \acp{fm}, presents challenges that traditional \comps are ill-equipped to handle. Static \comps, such as GCC~\citep{gcc2024} or LLVM~\citep{LLVM}, are designed with deterministic and well-defined source code in mind, whereas \ac{ai}-driven software demands continuous optimization and adaptation based on probabilistic reasoning, user feedback, and real-world data streams. This gap in adaptability is one of the core reasons why a new generation of \comps, such as \autocog, is required to handle the fluid and iterative nature of modern \ac{ai}-powered systems.

\subsection{Deep learning \comps}
\label{sec:background:deep-learning-compilers}

\Ac{dl} \comps have emerged as a response to the need for specialized optimization techniques that are tailored to the unique characteristics of \ac{dl} models. These \comps, such as TVM~\citep{chen2018tvm}, XLA~\citep{openxla2024}, and Glow~\citep{rotem2019glowgraphloweringcompiler}, focus on optimizing tensor operations, data flow, and memory usage to accelerate model inference and training. \ac{dl} \comps take advantage of hardware-specific instructions, including GPUs, TPUs, and other accelerators, to maximize throughput and reduce latency. Unlike traditional \comps, which focus on optimizing symbolic logic, \ac{dl} \comps are designed to optimize the execution of computational graphs~\citep{dean2012largescale}, handling the complexity of parallelization, operator fusion, and data movement with high efficiency.

However, despite their significant contributions to performance optimization in \ac{dl} workloads, \ac{dl} \comps are still relatively limited in scope when it comes to broader \ac{ai}-native software systems. They are primarily designed for the optimization of pre-defined neural network architectures, which is inadequate for the continuous, real-time optimization required in dynamic \fmware systems. Moreover, \ac{dl} \comps typically lack integration with higher-level abstractions like prompt engineering, reasoning modules, and multi-agent architectures that are integral to \ac{ai}-native systems. \autocog extends beyond these limitations by incorporating search-based techniques to optimize not only the deep learning components but the entire \fmware stack, including prompts, \acp{ca}, and agent-based frameworks, enabling real-time adaptation and evolution of \ac{ai}-native software.

\subsection{Prompt \comps}
\label{sec:background:prompt-compilers}


Automatic Prompt Engineer (\ape)~\cite{zhou2023large} frames the problem of automatically generating and selecting prompts as a ``natural language program synthesis'', drawing parallels between prompt engineering and traditional programming.
\ape uses an \ac{fm} to generate candidate prompt solutions, and a set of input-output demonstrations $\mathcal{D}_{\textit{train}}={(Q,A)}$ for solution evaluation (a.k.a. gold labels, see Section~\ref{sec:search-mechanism}).
\ape evaluates each candidate solution $\rho$ by prompting the \ac{fm} with the concatenation of $\rho$ and $Q$ and comparing the generated result with $A$ (see error estimator in Section~\ref{sec:search-mechanism}).
\ape can also be configured to run an Iterative Monte Carlo Search algorithm by applying a paraphrasing prompt (see heuristic approximator in Section~\ref{sec:search-mechanism}) to the candidate solutions and filtering out candidates with low scores.

\promptbreeder~\cite{fernando2024promptbreeder} uses an \ac{fm} to drive a ``self-improvement'' search process that iteratively mutates a set of candidate task-prompts and evaluates the fitness of the candidate task-prompts based on a ``training set'' (or gold labels).
One of the main innovations behind \promptbreeder is the idea of using ``self-referential'' mutation-prompts that are used to instruct the \ac{fm} to perform mutations over the task-prompts and also undergo mutation themselves.
In a similar vein, \evoprompt~\cite{guo2024connectingevoprompt} uses evolutionary algorithms to optimize a population of prompts by iteratively applying evolutionary operators (i.e., crossover and mutation) followed by the evaluation and selection of best-fit prompts to generate new offspring.

\protegi~\cite{pryzant2023automaticpromptoptimizationwithgradientdescent} simulates a ``gradient-descent'' approach to optimize prompt templates.
In a ``forward'' step, it uses a ``mini-batch'' of input data and a reflection prompt to generate ``gradients'', i.e., a summary, in natural language, of the associated ``error'' with the prompt under optimization for each of the instances of the mini-batch.
Afterwards, on the ``backpropagation'' step, a delta-prompt is used to edit the prompt under optimization towards the direction of the ``gradients'' (i.e., by observing the ``error'' summary generated in the ``forward'' step).
A beam search is then used to search over the space of candidate prompts.

\sammo~\cite{schnabel2024prompts} represents \fmware programs as ``symbolic prompt programs'', which are \acp{dag} with each node indicating an arbitrary function and an edge indicating a function call.
\sammo extends the search for optimal \fmware program configurations beyond prompt templates to include parameters of \promptware components.
While optimizing the \fmware program, \sammo uses metaprogramming to mutate the associated \ac{dag} (e.g., to change the format of a prompt template or remove a node from the computation graph).
Like the other approaches, \sammo uses labelled samples to evaluate the candidate solutions during the search procedure.
Search in \sammo can be either enumerative, where the search space is explicitly defined (e.g., for parameters of \promptware components), or iterative, where the search space is described implicitly by an initial state and a set of mutation operators (e.g., for prompt templates)
The mutation operators offered by \sammo can modify the prompt template text or the prompt template structure.

\section{Synthesizing Intents into \fmware}
\label{sec:fmware-synthesis}

Synthesizing human intents into fully functional \fmware involves more than just generating prompts. It requires integrating various \fmware components within a cohesive \ac{ca}. In the following, we introduce the concepts of \fmware (Section~\ref{sec:fmware-synthesis:fmware-concept}) and \acp{ca} (Section~\ref{sec:fmware-synthesis:cogarch}). Next, we discuss the role of \comps in the \ac{se30} era (Section~\ref{subsec:fmware-synthesis:role-compiler-se30-era}) and the challenges of synthesizing intents into \fmware (Section~\ref{sec:fmware-synthesis:intents-into-fmware}). Finally, we discuss the state of the practice (Section~\ref{sec:fmware-synthesis:compiler-sotp}) in \fmware \compilation.

\subsection{\fmware and related concepts}
\label{sec:fmware-synthesis:fmware-concept}

\begin{figure*}
\centering
\includegraphics[width=0.85\linewidth]{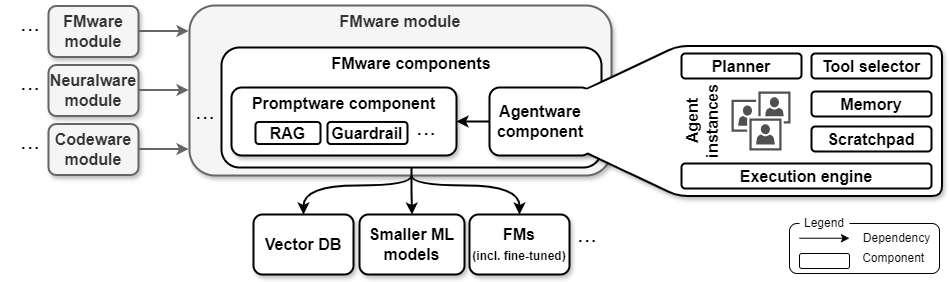}
\caption{Example of an \fmware module with one \fmware component comprised of a \promptware and an \agentware components.}
\label{fig:fmware_module}
\end{figure*}

\fmware is a software system that uses \acp{fm} like LLMs as fundamental building blocks~\cite{hassan2024rethinkingse}. An \fmware is further subdivided into two other categories: \promptware and \agentware. 
A \promptware is characterized by using prompts (typically written in natural language) to interact with the \acp{fm}, whereas \agentware is characterized by the usage of \acp{fm} to implement autonomous agents~\cite{park2023generativeagents,yao2023react} that are capable of decision-making and interaction with the environment or with other agents.
An \emph{\fmware program} is composed of one or more \emph{\fmware modules}, with each module sharing dependency relationships with \codeware (the software paradigm driven by logic-based source code), \neuralware (the software paradigm driven by feature-based and deep learning models), and other \fmware modules.
Each \fmware module comprises one or more \emph{\fmware components}, and each \fmware component comprises smaller, self-contained \promptware and \agentware components.
Figure~\ref{fig:fmware_module} depicts an \fmware module composed of a single \fmware component, which, in turn, comprises a \promptware and an \agentware component.

\subsection{\fmware and cognitive architectures} 
\label{sec:fmware-synthesis:cogarch}

A \ac{ca} defines the coordination mechanism that dictates \textit{what} software systems (possibly from different paradigms) will be invoked, \textit{how}, and \textit{in which order}. As we discuss in our prior work~\citep{hassan2024rethinkingse}, the choice of coordination mechanism is what differentiates \promptware from \agentware. In \promptware, coordination follows a static \textit{workflow} (or simply, \textit{flow}~\citep{mspromptflow}) that outlines a specific set of tasks and the sequence in which they must be performed. A task within this workflow can also trigger another workflow. On the other hand, \agentware employs a coordination mechanism driven by autonomous agents. These agents determine both the steps to be taken and the order in which they occur. In \acp{ca} that use multiple agents, coordination results from interactions between the agents themselves, such as in Microsoft Autogen~\citep{wu2023autogenenablingnextgenllm}. Patterns for \acp{ca} are starting to emerge~\citep{hassan2024rethinkingse}.

\subsection{The role of \comps in the \ac{se30} era}
\label{subsec:fmware-synthesis:role-compiler-se30-era}

\ac{se30} represents a significant shift towards an intent-first approach, where development is no longer driven by code but by \textit{intents} conveyed through interactive conversations between human developers and their \ac{ai} counterparts~\citep{hassan2024ainativesoftwareengineeringse}. We call this approach \textit{conversation-oriented development}. In \ac{se30}, \ac{ai} takes the lead in automating the code creation process by \textbf{synthesizing} intents into executable software. We note that the resulting software system might be \codeware, \neuralware, \promptware, \agentware, or any combination thereof. Due to the new use cases and business opportunities unlocked by \acp{fm} and generative \ac{ai}, this paper focuses on the synthesis of intents into \fmware. As such, the \comp must search for the most suitable \ac{ca} for the \fmware based on a set of competing criteria (e.g., accuracy, cost, and latency).

\subsection{The challenging task of synthesizing intents into \fmware} 
\label{sec:fmware-synthesis:intents-into-fmware}

\begin{figure*}
\includegraphics[width=0.85\linewidth]{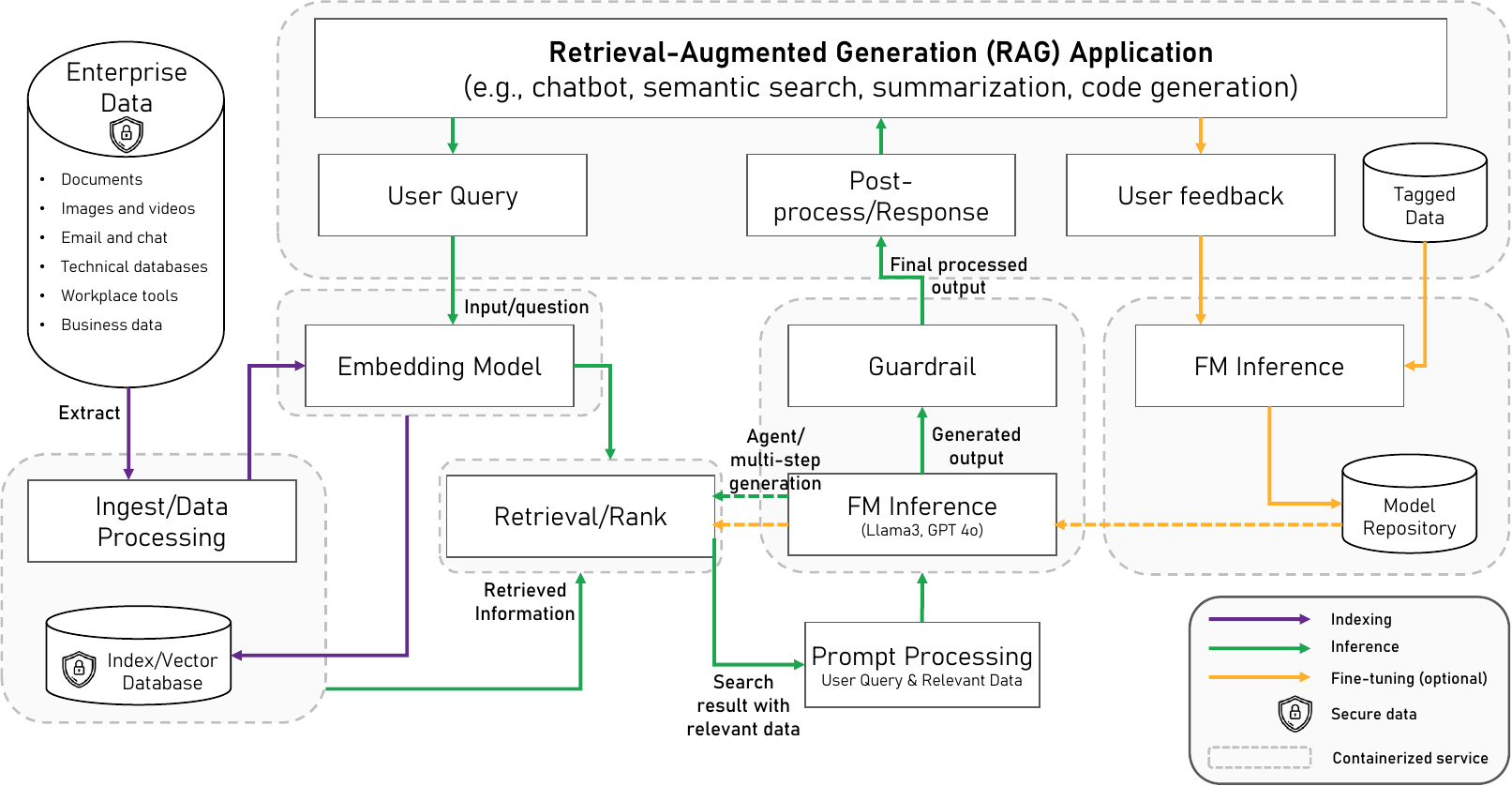}
\centering
\caption{RAG pipeline (adapted from OPEA's reference RAG pipeline~\citep{opea2024})}
\label{fig:rag_pipeline}
\end{figure*}

Synthesizing intents into \fmware is a very challenging task due to the complexity of \fmware. Gone were the days when \fmware was simply a thin wrapper around an \ac{fm} (e.g., GPT-4o). Analogously to how \ac{ml} models constitute only a minuscule part of \ac{ai} systems from the SE 2.0 era~\citep{Sculley15}, \acp{fm} also constitute an equally small part of the stack in the \ac{se30} era. Much of the code in an \fmware lies in (i) configuration (e.g., setting up \ac{fm} output parameters, such as temperature and top-k), (ii) data collection (e.g., curriculum engineering), (iii) RAG systems and embedding models, (iv) data verification (e.g., guardrails), (v) analysis tools (e.g., evals and benchmarks), (vi) process management (e.g., AIOps), (vii) machine resource management (e.g., Ray clusters), (viii) serving infrastructure (e.g., Ray serve), and (ix) monitoring (e.g., semantic observability). Due to the non-deterministic behavior of \acp{fm}, these components tend to be significantly more complex than those from \neuralware or \codeware. As an illustrative example, \ac{rag} itself further expands into a set of several interconnected components (see Figure~\ref{fig:rag_pipeline}).

We also emphasize that the many pieces of \fmware must evolve in response to feedback data, especially in the context of \textit{data flywheel} approaches~\citep{DataFlywheel}. The data flywheel is a self-reinforcing cycle in which data collection, analysis, and insights lead to ongoing improvements and growth within a system. As more data is accumulated (such as field, telemetry, or human feedback data), it enhances the accuracy and efficiency of algorithms and processes (e.g., model refinement and agent enhancements), which in turn improves user experiences and operational effectiveness. This attracts more users and generates additional activities, resulting in even more data, which drives further optimization. Consequently, \fmware remains in a continuous state of evolution.

\subsection{State of the practice}
\label{sec:fmware-synthesis:compiler-sotp}

\dspy~\cite{khattab2023dspy} is one of the first steps towards an end-to-end infrastructure to optimize an \fmware program, contributing programming \constructs to express a computation graph and an optimization method for the whole program.
\dspy's programming model~\cite{khattab2023dspy} represents an \fmware as a computation graph of \constructs that involve two abstractions: modules (to perform common prompt operations like reasoning procedures) and signatures (to define a module's input and expected output).
\dspy's compiler optimizes a \dspy program (written with the aforementioned programming model) by taking as input a ``training set'' (a.k.a. gold labels) and an optimization metric.
The compiler first generates candidate solutions (e.g., few-shot examples for a prompt template or specific control flows for the program) and then uses a search algorithm (e.g., random search) to find the best candidate solutions based on the given metric.

\textgrad~\cite{yuksekgonul2024textgradautomaticdifferentiationtext} adopts the same ``gradient-descent'' metaphor as \protegi (Section~\ref{sec:background:prompt-compilers}) to optimize an \fmware program described as a computation graph with a syntax that resembles that of the popular \ac{dl} \pytorch library.
In \textgrad, the ``gradients'' generated by the reflection prompt propagate through the computation graph, describing how each component of the \fmware program should be modified to improve the overall performance.
More specifically, each node $v$ of the computational graph represents unstructured data, such as natural language or an image. 
During compilation, these nodes perform a transformation on their associated data, following the aggregated ``gradient'' (e.g., a set of summary of ``errors'') from all successor nodes of $v$.
Objective functions in \textgrad are customizable and can be defined, for example, as a natural language text that is evaluated by an \ac{fm} or the output of running an \ac{fm}-generated piece of code against a set of test cases.
\textgrad also differentiates between instance optimization and prompt optimization. 
While the former refers to optimizing the solution to a problem, the latter refers to optimizing the prompt templates themselves.

Despite the importance of prior advances in the representation of FMware programs and prompt optimization, our vision for \autocog in the era of SE 3.0 extends beyond these foundations. Tools like DSPy represent significant and mature advances in FMware program optimization, providing programming abstractions and optimizers that realize key aspects of our vision. DSPy, in particular, supports program-level optimization of prompts, demonstrations, and module configurations, with pluggable optimizers and systematic approaches to improving FMware programs. Our vision for \autocog builds upon and extends these foundations in four complementary directions.

First, we envision integration with the SE lifecycle (detailed in Section~\ref{subsec:bridging-intent-compilation-with-se-practices}). While existing tools focus on optimizing FMware programs within their frameworks, \autocog addresses how intent compilation connects to requirements engineering, testing practices, CI/CD pipelines, maintenance and evolution, and collaborative development. This broader scope positions FMware compilation within complete software systems rather than as an isolated optimization problem.

Second, we propose a compiler infrastructure for cross-tool interoperability. This includes IRs that enable FMware programs written in different frameworks to share optimization advances, front-end/back-end separation allowing diverse FMware representations to use common optimizers, and community platforms for sharing compilation traces and gold-label datasets. While existing tools provide pluggable optimizers within their ecosystem, our vision extends toward a standardized compilation infrastructure that is agnostic to individual frameworks.

Third, \autocog explores multi-objective Pareto optimization rather than single-metric or weighted-combination approaches. Our prototype implements NSGA-II to simultaneously optimize competing objectives (e.g., accuracy, latency, cost) and present trade-off frontiers to users. While existing tools could optimize weighted combinations of metrics, true multi-objective optimization explores the Pareto frontier without requiring users to specify weights a priori.

Fourth, \autocog addresses compilation performance through systematic infrastructure (see Section 4.1). This includes semantic caching at multiple levels, compilation trace reuse that leverages prior compilations to accelerate future ones, distributed synthesis capabilities for parallelizing the search process, and transparent model serving optimization (including model layering and swapping). These performance optimizations are essential for making FMware compilation practical at scale, particularly in continuous integration environments where compilation time directly impacts development velocity.

We now turn to the architecture of \autocog, detailing the key components and design choices that realize these capabilities and support the research directions outlined in Section 5.

\subsection{Positioning SE 3.0 Compilers within Broader SE Paradigms}

While we frame our vision through the lens of compilation, we recognize that SE 3.0 compilers share conceptual similarities with other established top-down software engineering approaches. In particular, Model-Based Software Engineering (MBSE) has long explored the automatic generation of implementations from high-level specifications and models, and self-adaptive systems research has investigated mechanisms for runtime optimization based on quality objectives. Moreover, software product lines provide systematic approaches to managing variability through feature models and configuration.

We adopt the compiler framing for different reasons. First, compilation emphasizes systematic, repeatable transformation processes operating on well-defined representations, enabling rigorous optimization techniques. Second, the compiler analogy captures the distinction between compilation-time decisions (i.e., prompt synthesis and configuration selection) and runtime execution (e.g., FM invocation and agent coordination) of the FMware program. Third, compiler infrastructure naturally supports modular front-ends and back-ends, enabling interoperability across different FMware representations. Finally, the compiler abstraction provides a principled framework for reasoning about correctness, efficiency, and optimization in ways that complement but differ from model-driven or adaptive system perspectives. Yet, these are complementary rather than competing viewpoints. As we discuss in Section 5.2, SE 3.0 compiler research can benefit from techniques developed in MBSE, self-adaptive systems, and software product lines research.
    w\section{Compiler.next}
\label{sec:compiler-next}

\autocog represents a step forward in offering a solution to some of these challenges presented in our prior work discussion ten challenges for the engineering of trustworthy FMware~\cite{hassan2024rethinkingse}. In particular, our prior work outlines FMArts, a low-code, full lifecycle platform for engineering FMware comprised of three layers:

\begin{itemize}
    \item \textbf{FMware hub (top layer):} Contains reusable tools and infrastructure to support the FMware engineering lifecycle,
    \item \textbf{FMware framework:} Encompasses a set of conceptucal primitives to define and represent FMware programs,
    \item \textbf{Graph compiler and fusion runtime (bottom layer):} Provides infrastructure for running the FMware program optimized for the underlying FM deployment.
\end{itemize}

\begin{figure}[!htbp]
    \centering
    \includegraphics[width=0.45\linewidth]{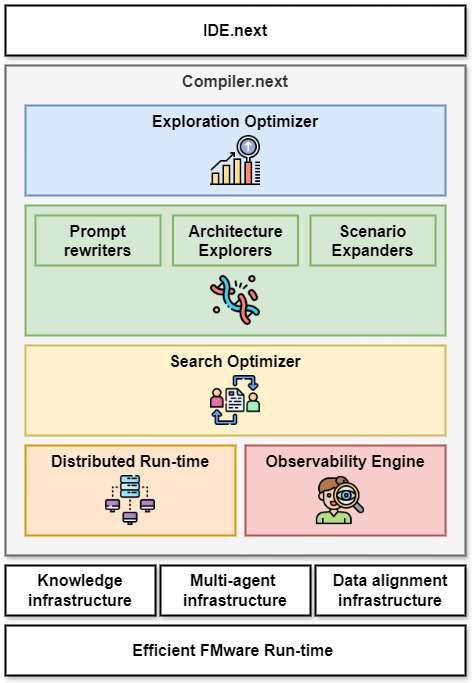}
    \caption{The technology stack of \autocog.}
    \label{fig:se30-stack}
\end{figure}

Within FMArts, \autocog plays the role of the higher-level program synthesizer that produces or refactors assets in the FMware Hub (e.g., prompts, agents, workflows) and provides a compiled plan to the Graph Compiler and Fusion Runtime for performance optimization and execution. Figure~\ref{fig:se30-stack} depicts the technology stack of \autocog. In the following, we discuss each of its components.

\smallskip \noindent \textbf{Cognition Exploration Optimizers.} Drives the search process efficiently and intelligently using techniques like self-reflection.
It also allows the specification of different optimization objectives to improve \fmware attributes such as correctness, latency, and cost. 

\smallskip \noindent \textbf{Prompt rewriters.} Focuses on enhancing and refining prompts, using advanced techniques to modify prompt structures for better outcomes.
A prompt rewriter can leverage the several prompt engineering techniques and prompt patterns that have been published in the literature (e.g., in the form of a curated library~\cite{schmidt2024promptcatalog}) to increase its chances of bootstrapping examples that will improve results.

\smallskip \noindent \textbf{Architecture explorers.} Searches for optimal configurations of RAG parameters (see Section~\ref{subsec:fmware-synthesis:role-compiler-se30-era}) and patterns of \acp{ca}, such as \ac{fm} Call (a basic architecture focusing on single function model calls), \ac{fm} Chain (a sequential model chain, where outputs of one model feed into the next), Router Agent (an architecture where an agent routes tasks based on the context or input type), State Machine Agent (a structured architecture that operates based on predefined states and transitions), and Autonomous Agents (fully autonomous agents that make decisions and act independently within the system).

\smallskip \noindent \textbf{Scenario expanders.} Expands the scenario description and cases used to drive the search by synthesizing new scenarios from existing ones that include new dimensions.
This ensures that, during the search, the \fmware was optimized for different target domains with a diverse set of scenarios. In the context of \autocog, a scenario is a structured representation of the task that the compiler seeks to solve during intent compilation. A scenario defines the problem space by specifying the user intent, the associated gold labels, and the evaluation criteria used to assess candidate solutions. Scenarios thus act as the optimization environment within which \autocog searches for the best configuration of prompts, cognitive architectures, and system parameters. \autocog’s scenario expander mutates these task descriptions by synthesizing new variations of a scenario from the existing ones. For example, given a code generation task with a particular docstring, the scenario expander may generate paraphrased versions of the docstring, alternative formulations of the input-output pairs, or additional test cases. This expansion increases the diversity of the optimization environment, encouraging \autocog to search for solutions that are robust across variations of the original task, reducing the risk of overfitting to narrowly defined task instances.

\smallskip \noindent \textbf{Search optimizer.} Leverages prior local and crowd-runs for more efficient driving of the search process.
In particular, uses the \compilation traces of prior \compilations as feedback information and reuses it to make the next \compilations more efficient (e.g., caching common \compilation steps).
Past search data can help in tuning the parameters of search algorithms (like genetic algorithms, simulated annealing, etc.) to better fit the problem space based on previous outcomes.
Insights from past searches can guide the development of new heuristics that are more adept at solving specific types of problems, thereby improving search efficacy.
Historical searches can also be used to understand user preferences, allowing for personalization.
By continuously incorporating insights gained from historical search data, systems can incrementally improve.

\smallskip \noindent \textbf{Distributed synthesizer run-time.} Uses a distributed platform to speed up the synthesizer. In the context of resource-limited devices (e.g., mobile), it is important that model layering or swapping can be transparently handled~\citep{RouteLLM,RouterBench}.

\smallskip \noindent \textbf{Synthesizer observability engine.} Enables debugging and traceability of the whole synthesizer, such that developers can understand the program states that caused issues and take action if needed. Debugging and traceability are needed at several levels of abstraction (e.g., \ac{ca} selection, optimizations within a given \ac{ca}, prompt optimizations).

\smallskip \noindent \textbf{Other layers.} On top of the stack is the \textit{IDE.next} component. \textit{IDE.next} is the new \ac{ai}-native IDE that powers software development in the \ac{se30} era~\citep{hassan2024ainativesoftwareengineeringse}. In turn, \autocog sits on top of different infrastructures that support \fmware implementation, such as \textit{knowledge infrastructure} to support \ac{rag}, \textit{multi-agent infrastructure} to drive autonomous agents, and \textit{data alignment infrastructure} to support \ac{fm} alignment (e.g., fine-tuning processes). Finally, the bottom layer involves serving \acp{fm} efficiently.

\subsection{The search mechanism}
\label{sec:search-mechanism}

Both \promptware and \agentware components have a combination of input parameters for which they perform optimally for a certain task.
For example, for a \promptware component, these parameters might be associated with the embedding models used in the \ac{rag} component or the parameters of the guardrail mechanism.
Similarly, for the \agentware components, parameters associated with the agents' \ac{ca} (such as the number of agents and the roles of each agent) and memory components (such as the memory buffer size) need to be determined.
At the same time, the prompt templates used to interface the \fmware components with the \acp{fm} also impact the general performance of the \fmware application.
Therefore, there is a need to find the set of components' parameters and prompt templates that lead to the best performance of the \fmware application, while the search space of such parameters is too large to be explored manually.

\begin{figure*}
\centering
\includegraphics[width=0.65\linewidth]{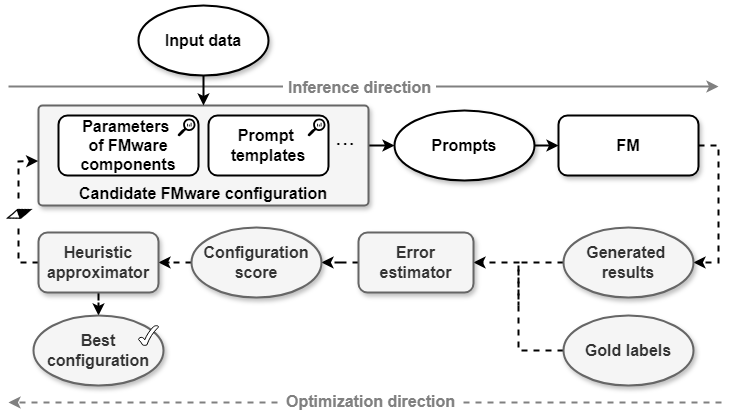}
\caption{Search iteration steps in \autocog.}
\label{fig:fmware_opt}
\end{figure*}

Figure~\ref{fig:fmware_opt} depicts the general search mechanism of \autocog, highlighting the continuous optimization of components' parameters and prompt templates.
The optimization process starts by instantiating the optimizable components of an \fmware and inputting task-dependent data to these components, generating a specific \emph{\fmware configuration} based on the instantiated components.
In the example of Figure~\ref{fig:fmware_opt}, we focus on the set of components' parameters and prompt templates of \fmware applications, but other components can also be subject to optimization.
Afterwards, an inference step is performed by passing prompts to the target \ac{fm}, generating results for the specific combination of prompts and \ac{fm}.
The \emph{error estimator} then compares the generated results against a set of target \emph{gold labels} that describe expected results for the specific task, generating a \emph{configuration score} that estimates how close the generated results are to the optimal solutions (based on the gold labels as a reference of optimality).
The error estimator can adopt different approaches to calculate the configuration score as long as they are suitable for the task performed by the \fmware component under optimization.
For example, to compare the generated results against the gold labels, a code generation task can use either text-based metrics, such as BLEU~\cite{aryaz2024crystalbleu}, or test-based metrics, such as computational accuracy~\cite{pan2024lostintranslation}.
In the current design of \autocog, we assume that gold labels are provided as static inputs to the search process. However, we recognize that in practice, specifications often evolve during compilation. For instance, requirements may be refined as developers gain a better understanding of the task, or new evaluation cases may be generated as the system exposes unanticipated behaviours.
Finally, a \emph{heuristic approximator} records the best configuration found during the optimization process and applies operations to modify the components' parameters and prompt templates to generate new candidates of \fmware configuration and steer the optimization process towards better candidate solutions.
The heuristic approximator can use different methods to modify the components' parameters (e.g., random search~\cite{bergstra2012randomsearch}) and prompt templates (e.g., using a \ac{fm} to rewrite the template or using differentiable prompt formats such as soft prompts~\cite{wu2023infopromptsoftprompt}).
The optimization process is repeated until a certain stop criterion is achieved, such as the number of iterations, \compilation time, or a threshold configuration score.


\smallskip \noindent \textbf{Conceptual model.} In \autocog, each component of an \fmware program is represented by an \class{Operation} that implements an arbitrary functionality and contains both static and dynamic parameters, with some of the dynamic parameters being optimizable.
For example, an \class{Operation} can represent a \promptware component that performs a \ac{fm} inference, for which a prompt template can be optimized, or retrieval from a vector database (\ac{rag}), for which the parameter associated with the top-$k$ documents can be optimized.
All \class{Operation} elements have a common interface that can be used to specify optimizable parameters and an \class{Optimizer}.
An \class{Operation} is a stateful entity that can exchange messages with another \class{Operation}, being suitable to different \fmware program representations, such as prompt chains~\cite{Langchain-Ai}, computation \acp{dag}~\cite{schnabel2024prompts,yuksekgonul2024textgradautomaticdifferentiationtext}, or \acp{ca}~\cite{hu2024automateddesignagenticsystems}.

An \class{Optimizer} in \autocog is a pluggable component that specifies how the parameters of an \class{Operation} are optimized.
For instance, one of the optimizers implemented in \autocog uses an \nsgii multi-objective \ac{ga}~\cite{deb2002nsgaii,blank2020pymoo} to optimize prompt templates, allowing the customization of all genetic operators such as crossover, and mutation~\cite{deb2002nsgaii} in addition to the objective functions.
This prompt template optimizer uses a \ac{fm} to drive the search (e.g., to mutate a prompt candidate) and receives as parameters the associated intent with the \class{Operation} (e.g., ``generate source code from documentation''), the input data (e.g., the function signature and documentation), the gold labels (e.g., unit tests), and references to both the \emph{evaluator-} and \emph{release-\acp{fm}} used to drive the search and evaluate the candidate results, respectively.
An \class{Optimizer} can be extended or customized according to the task performed by the associated \class{Operation}.
For instance, for the optimization of enumerable component parameters, our \ac{ga}-based optimizer can use some genetic representation of the parameters (e.g., an array of binaries for integers and floats)~\cite{eiben2003introtoevolcomputing}, avoiding the high costs of using a \ac{fm} to drive the search.

An \class{Optimizer} also has an aggregation relationship with another customizable element of the \autocog framework called \class{EvaluationBench}, which defines the format and evaluation logic of the gold labels used during the optimization process.
This is an important aspect of the framework because gold labels highly depend on the task performed by the associated \class{Operation} with an \class{Optimizer}.
In \autocog, we also separate the \acp{fm} used for application deployment (release-\ac{fm}) from the \ac{fm} used to drive the search (evaluator-\ac{fm}) because the former can be a smaller \ac{fm} that does not have the same capabilities of the latter to generate diverse and quality candidate solutions during the optimization process.

\subsection{Illustrative Example: Optimizing Prompts for Code Generation}

To illustrate the working mechanism of \autocog, we use a running example of a code generation FMware containing a single Promptware component. The FMware receives as input a function signature with documentation describing the desired functionality and generates the implementation. The compilation objective is to synthesize an optimized system prompt template that maximizes accuracy and performance on code generation tasks (i.e., generating semantically correct code that runs fast). The software maker provides \autocog with an intent (``Generate source code from documentation''), a collection of problems (function signatures with documentation), and gold labels consisting of unit tests for evaluation.

\begin{figure}
    \centering
    \includegraphics[width=\textwidth]{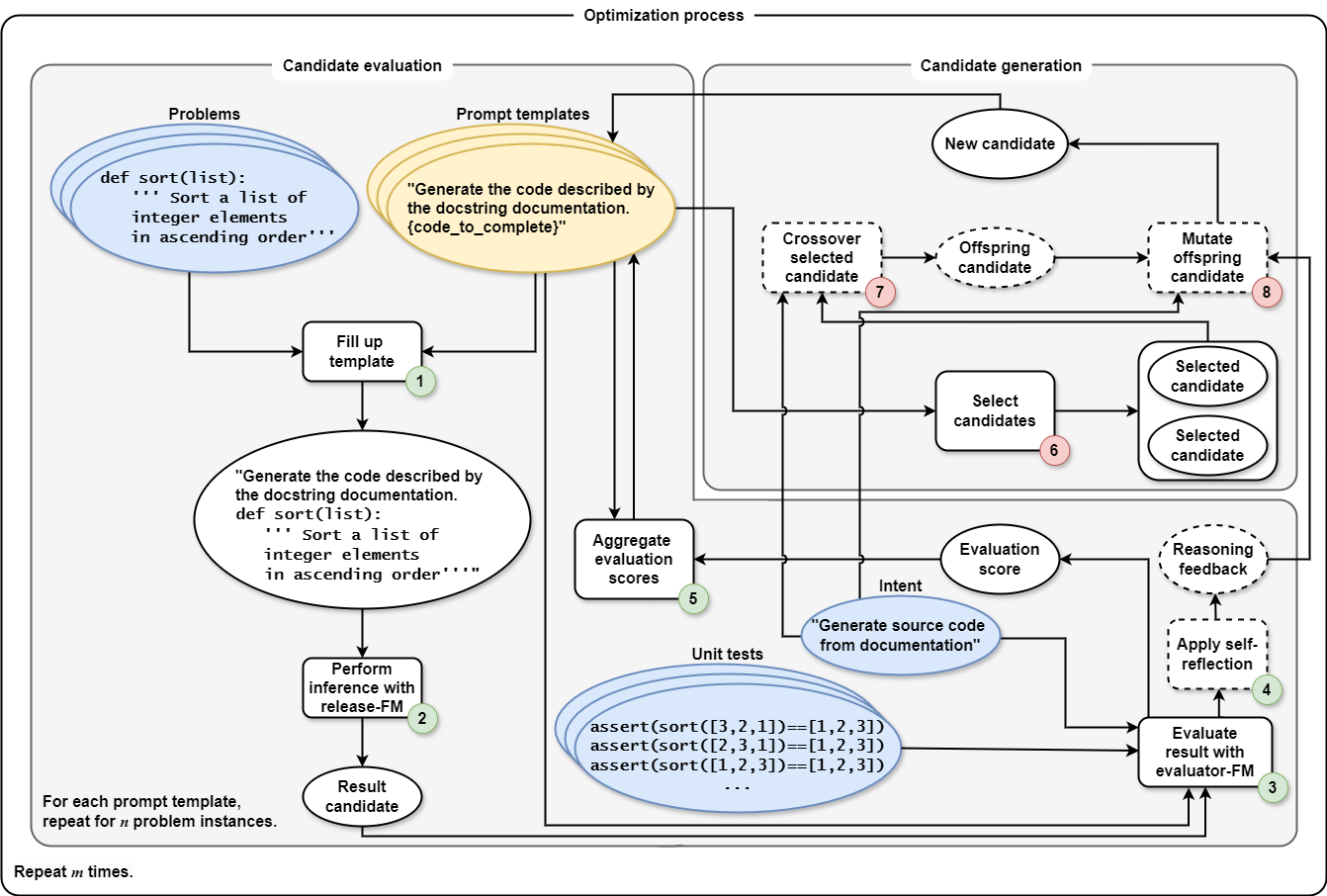}
    \caption{The operation of one of \autocog's optimizers in our running example of an \fmware for code generation. The dashed box indicates optional steps.}
    \label{fig:running-example}
\end{figure}

Figure~\ref{fig:running-example} illustrates the optimization process, which operates through iterative cycles of candidate evaluation and candidate generation. The process begins with an initial population of prompt template candidates, either user-provided or automatically generated by an FM based on the intent, and refines this population over multiple iterations to identify high-performing configurations.

\noindent \textbf{1. Fill up template:} The first step of the optimization occurs during the candidate evaluation process (left side of Figure~\ref{fig:running-example}). Each prompt template contains placeholders that are instantiated with information from a problem instance. In our example, the placeholder \prompt{code\_to\_complete} is replaced with a function signature and its documentation string. During optimization, the placeholder content remains static, while other template components (e.g., few-shot examples, instructions) are optimized. Each component can be optimized separately. The results of template instantiation are cached, as the same problem-template pairs frequently recur across iterations.

\noindent \textbf{2. Perform inference with release-FM:} The instantiated prompt is executed using the release-FM to generate a result candidate (in this case, completed code that concatenates the function signature and documentation with the FM-generated implementation). The prompt template is optimized specifically for the FM that will be deployed in the production FMware application. All FM inference calls are cached based on their input prompts, substantially reducing costs when identical or similar prompts are evaluated in subsequent iterations.

\noindent \textbf{3. Evaluate result with evaluator-FM:} The result candidate is evaluated using two objective functions: (1) computational accuracy, measured as the proportion of unit tests that pass when executed against the generated code, and (2) execution latency, measured as the time required to execute the generated code. The evaluation approach is customizable and can incorporate additional quality attributes (e.g., code complexity) depending on user preferences and downstream requirements. Evaluation results are likewise cached, enabling rapid re-assessment of previously encountered candidates.

\noindent \textbf{4. Apply self-reflection:} Optionally, when unit tests fail, self-reflection prompts~\cite{shinn2023reflexion} can generate reasoning feedback about how the prompt template should be modified to improve performance. A key challenge with this technique is that feedback often becomes overly specific to individual problem instances. Since the optimization objective is to maximize performance across all problem instances, including out-of-sample instances at runtime, measures must ensure the feedback remains sufficiently general. The evaluator-FM, which may differ from the release-FM, produces this self-reflection feedback to leverage superior emergent abilities in generating actionable guidance.

\noindent \textbf{5. Aggregate evaluation scores:} Each prompt template is evaluated against $n$ different problem instances, and the individual scores are aggregated to compute an overall fitness score for the template. This aggregated score quantifies how effectively the template enables the release-FM to generate correct code across diverse scenarios.

\noindent \textbf{6. Select candidates:} During the candidate generation process (right side of Figure~\ref{fig:running-example}), the optimizer selects prompt templates based on their aggregated evaluation scores. The selection algorithm employed in our NSGA-II implementation uses crowding distance to select dispersed individuals on the Pareto front~\cite{deb2002nsgaii}, though alternative approaches such as roulette-based selection can be configured to foster diversity during search space exploration~\cite{jebari2013selection}.

\noindent \textbf{7. Crossover selected candidates:} Selected templates undergo crossover, which generates offspring by instructing an FM to combine elements from parent templates while considering the provided intent. As a customizable operator, crossover can adopt lower-cost approaches that avoid FM inference or can be skipped entirely. The results of crossover operations are cached to avoid regenerating identical offspring in subsequent iterations.

\noindent \textbf{8. Mutate offspring candidate:} Offspring templates are mutated by instructing an FM to modify the prompt based on the intent. To reduce costs, mutation can combine FM inference with simpler token replacement strategies~\cite{taherkhani2024epiccosteffectivesearchbasedprompt}. Mutation results are also cached, as identical mutations may be proposed multiple times as the population converges. The newly generated candidates enter the evaluation phase in the subsequent iteration, completing the optimization cycle.

The flexibility of \autocog enables implementation of diverse search strategies. For instance, a gradient descent-like approach can be implemented using a crossover operator that forwards selected candidates to a mutator that modifies templates based on self-reflection feedback. Alternatively, \autocog can evaluate a set of prompt template alternatives and select the best performer, generating reasoning traces about potential improvements. \autocog also leverages parallelization by evaluating multiple candidate results concurrently, further reducing overall compilation time.

As iterations progress, the population gravitates toward configurations that consistently achieve high evaluation scores on the gold labels. Upon convergence, \autocog selects the highest-performing prompt template from the final population as the optimized FMware artifact, ready for deployment in the code generation assistant. This example demonstrates how \autocog applies systematic, search-based optimization to synthesize effective FMware configurations, transforming an abstract intent into a concrete, empirically validated prompt template through automated exploration of the design space.

\subsubsection{Cost optimization through semantic caching}
The iterative nature of the optimization process creates substantial opportunities for cost reduction through semantic caching. Unlike traditional exact-match caching, \autocog employs semantic caching that stores embeddings of prompts and retrieves cached results for semantically similar inputs. For FM inference calls, both to the release-FM during code generation and to the evaluator-FM during assessment, and for genetic operator outputs (crossover and mutation), the cache compares the semantic similarity of the current input against previously cached entries using a configurable similarity threshold.

This similarity threshold introduces a critical trade-off between compilation speed and the exploration of the search space. A higher threshold (e.g., 0.95) requires near-exact semantic matches, resulting in fewer cache hits but ensuring diverse candidate evaluation across the search space. Conversely, a lower threshold (e.g., 0.80) increases cache hit rates and dramatically reduces compilation time and FM invocation costs, but risks limiting exploration by treating semantically similar yet potentially distinct candidates as equivalent. The optimal threshold depends on the compilation scenario: early iterations may benefit from lower thresholds to quickly eliminate poor regions of the search space, while later iterations may require higher thresholds to distinguish between high-quality candidates with subtle differences.

The semantic caching mechanisms prove particularly effective as the population converges toward high-quality regions of the search space, where candidates increasingly share structural components and generate similar prompts. Additionally, each iteration of the optimization loop is checkpointed, enabling recovery from failures without repeating expensive computations. The cumulative effect of semantic caching, when properly tuned, substantially reduces both compilation time and the number of costly FM invocations required while maintaining sufficient search diversity to discover high-performing configurations.

\subsubsection{Co-optimizing composed FMware components}
The example presented above focuses on optimizing a single Promptware component in isolation. However, FMware applications often comprise multiple composed components with dependencies between them. For instance, an Agentware component may contain multiple Promptware components, each with its own prompt template, or a RAG component may precede a Promptware component in a processing pipeline. Our current implementation of \autocog addresses this compositional challenge through joint optimization, where the search space encompasses configurations across multiple components simultaneously, evaluating candidates based on the end-to-end performance of the composed FMware. In this approach, the optimizer explores parameter combinations across all components (e.g., RAG retrieval settings and prompt template structures) to identify configurations that maximize overall system performance.

An alternative approach for SE 3.0 compilers is hierarchical optimization, in which components are optimized in stages, with information flowing between phases. For example, in a code generation system where a RAG component retrieves relevant code examples before the Promptware component generates the completion, after optimizing the prompt template, the compiler could leverage this information (such as the structure of the optimized prompt, performance signals indicating how retrieval quality affects downstream generation, or self-reflection feedback) to guide the optimization of the RAG component's parameters (e.g., the number of examples retrieved, the similarity threshold, or the embedding model used for retrieval). A subsequent optimization pass of the Promptware component could then refine the prompt given the updated RAG configuration. This sequential strategy offers potential advantages, including controlled search space dimensionality and opportunities for pipelining (e.g., optimizing one component while another is being evaluated), though it may miss synergies between component configurations that joint optimization is better at discovering. The compositional nature of FMware architectures and the trade-offs between these optimization strategies represent important directions for future SE 3.0 compiler research.

\subsection{Initial validation}

To provide a concrete proof of concept of \autocog, we conducted a controlled case study using the HumanEval-Plus benchmark, a widely used dataset for evaluating code generation systems. HumanEval-Plus consists of Python programming tasks where the goal is to generate the body of a function given its signature and docstring. This case study focuses on two central features of \autocog. First, we demonstrate the automated synthesis of system prompts from user-provided intents. Starting from the user-provided intent (``Generate source code for a function given a problem description found in the function’s documentation''), \autocog generates candidate prompts without human intervention. The search mechanism used by the system prompt synthesizer leverages the NSGA-II multi-objective evolutionary algorithm to optimize the candidates across three objectives: accuracy, latency, and execution cost (see Figure~\ref{fig:fmware_opt}). Accuracy is measured as the proportion of generated solutions that pass all the unit tests associated with each HumanEval-Plus entry. Latency corresponds to the runtime required to evaluate a candidate solution. Execution cost is measured by the number of tokens consumed per run (input and output). This setup demonstrates how intent-level specifications can be automatically compiled into optimized system prompts while balancing trade-offs between competing objectives.

Second, we evaluate the caching mechanism that accelerates the compilation process by reusing results of previously executed operations. Specifically, we implemented a tracer that manages a two-level semantic cache (we use an Euclidean similarity threshold of 0.85). At the first level (L1), cache entries are indexed by a hashable representation of the combination of the intent and the description of the training data, with each entry pointing to an associated second level (L2) cache. At L2, cache entries are indexed by a hash of the optimizable parameter (in this case, the prompt under evaluation). Each L2 entry stores the result of a time-consuming operation, such as a fitness calculation, crossover, or mutation in our NSGA-II implementation of the optimizer. When the optimizer requires one of these operations, the tracer intercepts the request and then attempts to locate a matching entry in the L1 cache. On a hit, it proceeds to the corresponding L2 cache, where a successful match immediately returns the stored result, such as the fitness score of a previously evaluated prompt. If a miss occurs at either level, the operation is executed and its result is stored in the relevant L1 and L2 cache entries. This design reduces redundant computations while ensuring that the optimization process remains faithful to the intended search dynamics.

In our evaluation, we allowed five generations of ten candidate solutions per task, resulting in a total of fifty candidates per optimization run. Seventy percent of the HumanEval-Plus tasks were held out as gold labels to guide the optimization process, while the remaining thirty percent were used for evaluation. We tested two foundation models, Qwen2.5-7B-Instruct and GPT-4o-mini, and measured performance for the initial synthesized prompts as well as the best optimized prompts after five generations. We further compared runs using GPT-4o-mini with caching enabled against runs without caching to highlight the trade-off between compilation efficiency and diversity of search space exploration. The evaluation started with a cold cache.

The results, summarized in Table~\ref{tab:humaneval}, show that \autocog improves accuracy while also reducing latency and execution cost across both models. Table~\ref{tab:cache} shows that the caching mechanism provides speed-ups of 22.1\% by avoiding redundant evaluations, though at the cost of a reduction in exploration, as previously seen solutions are favoured over newly generated ones. Overall, this controlled case study demonstrates the feasibility of \autocog and some research opportunities for optimizing intent compilation for AI-native software engineering.

\begin{table}[ht]
\centering
\caption{Evaluation of \autocog on the HumanEval-Plus benchmark (30\% held-out). We report accuracy (\%), average latency (s), and average number of tokens used per run for the best initial synthesized prompt and the best optimized prompt.}
\label{tab:humaneval}
\begin{tabular}{llrrr}
\toprule
\textbf{Model} & \textbf{Metric} & \textbf{Initial} & \textbf{Optimized} & \textbf{Improvement (\%)} \\
\midrule
\multirow[t]{3}{*}{Qwen2.5-7B-Instruct} 
  & Accuracy       & 0.26 & 0.56 & 46.4 \\
  & Avg. latency (s)    & 14.2  & 10.8  & 76.6 \\
  & Avg. \# tokens     & 537.1  & 369.3  & 68.7 \\
\midrule
\multirow[t]{3}{*}{GPT-4o-mini} 
  & Accuracy (\%)       & 0.68 & 1.00 & 47.0 \\
  & Avg. latency (s)    & 8.7  & 5.0  & 42.5 \\
  & Avg. tokens (\#)   & 500.0 & 417.1  & 16.5 \\
\bottomrule
\end{tabular}
\end{table}

\begin{table}[ht]
\centering
\caption{Comparison of \autocog performance with and without caching on the HumanEval-Plus benchmark using GPT-4o-mini. We report accuracy (\%), average latency (s), and average number of tokens used per run for the best optimized prompt. Total time is the end-to-end compilation runtime.}

\label{tab:cache}
\begin{tabular}{lrrr}
\toprule
\textbf{Metric} & \textbf{Without Cache} & \textbf{With Cache} & \textbf{Difference} \\
\midrule
Accuracy           & 1.00    & 0.70    & -30\% \\
Avg. latency       & 5.0     & 5.9     & 18\% \\
Avg. \# tokens     & 417.1   & 467.0   & 12\%  \\
Total running time & 10m:27s & 8m:15s  & -22.1\% \\
\bottomrule
\end{tabular}
\end{table}

The design of \autocog not only defines the architectural principles of FMware compilation but also directly surfaces a set of open research challenges. In the next section, we examine these challenges in detail, showing how each derives from the framework’s features and illustrating how \autocog can guide and structure future research on FMware.
\section{Research \& Development Roadmap for Compiler.next}
\label{sec:calls-for-action}

In this section, we discuss research challenges and opportunities for advancing intent compilation in SE 3.0. In Section~\ref{subsec:calls-for-action}, we discuss 10 calls for action that express our vision of the core SE challenges to be solved around this topic. In Section~\ref{subsec:research-opportunities-topdown-se}, we discuss how \autocog and our calls for action intersect with ``top-down'' SE methods like model-based software engineering (MBSE) and self-adaptive systems, inspiring researchers to explore these intersections to solve SE3.0 compilation challenges. Finally, in Section~\ref{subsec:bridging-intent-compilation-with-se-practices}, we discuss how our calls for action connect to well-established SE practices and the core research questions around these connections.

\subsection{Calls for action}
\label{subsec:calls-for-action}

This section describes a \ac{randd} roadmap that culminates in \ncfas \challenges that express our view on the \ac{se} challenges around the implementation of \ac{se30} \comps.
These challenges were derived based on a combination of our experience developing a prototype of \autocog, in-depth literature reviews~\cite{li2024seandfmindustryblog}, and discussions with top academics and industry leaders during several events (e.g., FM+SE Vision 2030, FM+SE Summit 2024). 
Building on the \autocog framework described in Sections 3 and 4, we organize the \ncfas calls for action of FMware research into four dimensions and associate each call for action with concrete \autocog features. Table~\ref{tab:dimensions} summarizes these dimensions, showing how the framework defines the architecture of FMware compilation and structures the \ac{randd} agenda we propose.
The next subsections present each of the \challenges.
We map each \challenge to the associated component of the technology stack shown in Figure~\ref{fig:se30-stack} by matching the icons of the figure and the section titles.
Subtitles with the \includegraphics[height=\baselineskip]{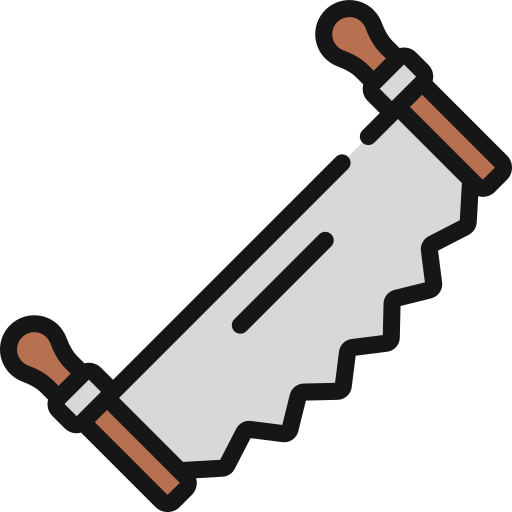} icon indicate cross-cutting \challenges.

\bgroup
\def\arraystretch{1.5}%
\begin{table}[ht]
\centering
\caption{Dimensions of \autocog and their relationship to the Calls for Action.}
\label{tab:dimensions}
\begin{tabular}{p{0.15\linewidth}p{0.3\linewidth}p{0.15\linewidth}p{0.3\linewidth}}
\toprule

\textbf{Dimension} & \textbf{\autocog's design feature} & \textbf{Calls for Action} & \textbf{Rationale} \\
\midrule

FMware program representation & 
- Representation of FMware modules as modular compositions of parameterized heterogeneous artifacts (Sec.~\ref{sec:fmware-synthesis:fmware-concept}).
\newline
- Architecture explorers search for optimal configurations of parameters of heterogeneous artifacts (Sec.~\ref {sec:compiler-next}).
&
Calls 1, 2, and 9 &
By composing FMware from modular components and supporting the optimization of multiple architectural patterns, SE 3.0 compilers motivate the need for programming constructs, heterogeneous compilation, and interoperability to be useful for a range of applications written with different technology stacks. \\

Result Validation & 
- Synthesizer observability engine enables debugging and traceability of the whole synthesizer and its outputs (Sec.~\ref{sec:compiler-next}), including how quality attributes of the generated program change during compilation.
\newline
- Compilation traces involving successful trajectories are re-used in future compilations, while failed trajectories can be used as feedback information to understand common error patterns (Sec.~\ref{sec:compiler-next}). &
Calls 5 and 7 &
Observability and trace reuse support both runtime validation and reproducible builds, ensuring FMware trustworthiness. \\

Computational Performance & 
- Caching common compilation steps to make the next compilations more efficient, with results from past searches guiding the development of new heuristics (Sec.~\ref{sec:compiler-next}).
\newline
- Distributed synthesizer run-time uses a distributed platform to speed up the synthesizer, including transparent handling of model layering or swapping (Sec.~\ref{sec:compiler-next}).
&
Calls 3, 6, and 10 &
SE 3.0 compilers achieve performance by combining heuristics, distributed execution, and trace reuse. These characteristics motivate research into efficient heuristics and community trace sharing. \\

User Priorities \& Goals & 
- Cognition exploration optimizers allow the specification of different user-defined and context-dependent optimization objectives (Sec.~\ref{sec:compiler-next}).
\newline
- Gold labels nudge the optimization to improve the FMware in specific instances and use cases that are of interest to the users, and can automatically evolve during compilation.
 &
Calls 4 and 8 &
SE 3.0 compilers must adapt to user goals, balancing multiple optimization objectives with a broad evaluation of user-prioritized scenarios.\\
\bottomrule
\end{tabular}
\end{table}
\egroup


\subsubsection{Quality \constructs for \fmware representation  \includegraphics[height=\baselineskip]{figs/crosscut-saw.png}}
\label{subsec:cfa:fmware-representation}

The development of \ac{se30} \comps would benefit from establishing \constructs with semantics that can represent all common operations and control logic of an \fmware program and can be translated to a \ac{ir} that different back-end optimizers can manipulate.
Intermediate representations are a foundational concept in compiler design, serving as the bridge between source languages and target architectures. A well-designed IR must be accurate—capable of representing source code without loss of information—and independent of any particular source or target language. The use of IRs has enabled compiler systems like LLVM to support many different source languages, generating code for many different target architectures~\cite{LLVM}, demonstrating the power of this abstraction for compiler modularity and reusability.

Modern compiler infrastructures have evolved toward increasingly flexible IR designs. LLVM's intermediate representation provides a language-independent, typed assembly language that serves as a portable format for optimization across multiple passes. More recently, MLIR (Multi-Level Intermediate Representation) has extended this concept by allowing multiple levels of abstraction to coexist within the same compiler system, with dialects enabling domain-specific operations while maintaining interoperability. This multi-level approach has proven particularly valuable for heterogeneous computing scenarios and domain-specific optimization.

For SE 3.0 compilers, such \constructs should allow the definition of 
intents~\cite{levi2024intentbasedpromptcalibrationenhancing} while avoiding eventual ambiguities, which might be particularly challenging if the intents are expressed using natural language.
We also argue that \comps should be capable of \compiling existing \fmware programs written in any representation (e.g., \langchain and \autogen) by providing a front-end component that can process programs written in such existing representations, transform them into this \ac{ir}, and send them to a back-end \compilation process whenever needed.

As an analogy with \ac{dl} compilers that use tensor-oriented \constructs (tensor in, tensor out) as building blocks of the computations, \ac{se30} \comps can adopt prompt-oriented \constructs (prompt in, generation out) as such building blocks.
Such \constructs should be attentive to common \ac{pl} quality criteria such as readability, orthogonality, and simplicity.
In addition, the definition of proper \constructs to represent the control logic of \agentware, in particular, is still an open problem, as it is in a continuum spectrum between fully manual -- where software makers specify all the execution paths based on the evaluation of inputs and outputs -- and fully automated -- where the \ac{fm} drives all decisions based on reasoning techniques such as \ac{cot}~\cite{wang2023selfconsistency}.

\cfa{\fmware representation needs \constructs with semantics that allow software makers to express complex \fmware programs through intentions. In particular, such \constructs should be able to represent every component of an \fmware and their interactions, with different levels of controllability, abstraction, and programmability. Such \constructs can then be transformed to a common \ac{ir} that can interoperate with different backend optimizers.}


\subsubsection{End-to-end \fmware optimization  \includegraphics[height=\baselineskip]{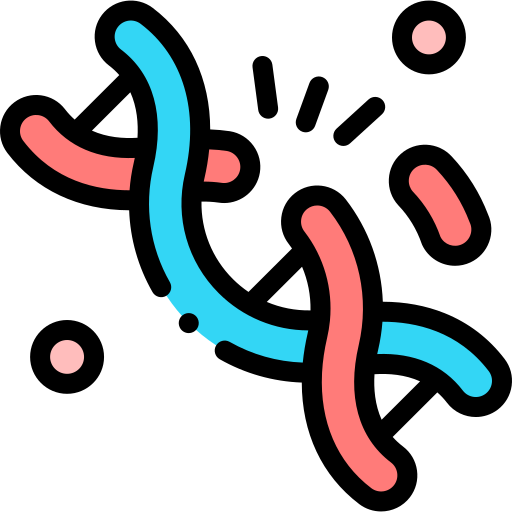}}

An \fmware is built with several interacting components connected to different \acp{fm}.
To interface with the \acp{fm}, these components generate dynamic outputs that are passed as arguments to prompt templates, which in turn has a significant impact on the outputs generated by the \acp{fm}~\cite{anwar2024foundational}.
As such, fully automating the search for the optimal \fmware configuration goes beyond the simple definition of prompt templates and also requires optimizing the parameters of all system components that interact with a \ac{fm}.
For example, an \fmware that applies a \ac{rag} pipeline must determine the best combination of encoder, retriever, and generator~\cite{lewis2021rag}, as the retrieved information will the part of the prompt given to a \ac{fm}. 
Similarly, a multi-agent component of an \fmware requires the determination of the agents' data and control flow, the \ac{ca}~\cite{romero2023synergistic,sumers2024cognitive} and the different reasoning strategies~\cite{josifoski2024flowsbuildingblocksreasoning} that the agents carry out to perform sophisticated computational processes with the \ac{fm}~\cite{hu2024automateddesignagenticsystems}.
Therefore, if the \compilation process aims at optimizing the overall performance of the \fmware, then all the system parameters deserve the attention of the optimization process.

\cfa{The \compilation of \fmware needs to move beyond the optimization of prompt templates in isolation to include the co-optimization of the different components of the \fmware. Despite the recent developments towards this direction, all \fmware components must be considered, particularly with better support for \agentware-based applications.}

\subsubsection{Effective search heuristics  \includegraphics[height=\baselineskip]{figs/gene-mutation.png}}

\Compiling an \fmware program involves the joint optimization of prompt templates and several other system parameters.
Therefore, the search space for \compiling an \fmware considerably increases as the number of components and their parameters increases, making the \compilation an intractable problem requiring a heuristic search.
Search-Based Software Engineering (SBSE) has established a rich repertoire of optimization techniques for SE problems~\cite{harman2012search,afzal2009asystematic}, including genetic algorithms, simulated annealing, and constraint-solving approaches. These techniques have been successfully applied to diverse problems such as test case generation, requirements selection, and refactoring~\cite{mariani2017asystematic}. More recently, program synthesis research has explored both classical search methods and neural approaches for generating code from specifications~\cite{gulwani2017program,kant2018recent}, demonstrating that hybrid techniques combining search with learned heuristics can be particularly effective.

However, synthesizing FMware presents distinct challenges that differentiate it from traditional program synthesis. While classical program synthesis operates on well-defined formal specifications and deterministic execution semantics, FMware compilation must navigate the probabilistic nature of FMs, the semantic rather than syntactic correctness criteria, and the need to co-optimize heterogeneous components (prompts, RAG parameters, agent configurations) simultaneously. Moreover, searching in the FMware domain often involves the repeated and systematic application of \ac{fm}-driven heuristics (e.g., prompt mutations) followed by (potentially) expensive evaluations of the intermediate solutions, which calls for an evidence-based determination of cost-effective heuristics that guide the search to quickly converge to a near-optimal solution.

Despite the extensive literature on search heuristics, it is not always clear which properties of a prompt or which parameters of each \fmware component contribute the most to the overall performance of an \fmware application.
For example, the large majority of the empirical research that examined the impact of changing a prompt in the \ac{fm}'s response focuses on the \emph{structure} of the prompt~\cite{cao2024worstpromptperformancelarge,lu2022fantasticallyorderedprompts,sclar2024quantifying,errica2024didiwrongquantifying}, but other properties of the prompt (and of an \fmware parametrization, more generally) still need research.
Understanding these properties would enable the adaptation of established SBSE techniques (such as informed mutation operators~\cite{hanna2025reinforcement} and fitness landscape analysis~\cite{sahin2026causes}) to the unique characteristics of FMware optimization.

\cfa{Empirical research is needed to demonstrate the prompt features and \fmware parameters that influence \ac{fm}'s output the most, such that search heuristics are defined based on such features and the search budget can be spent more effectively.}

\subsubsection{Gold labels construction  \includegraphics[height=\baselineskip]{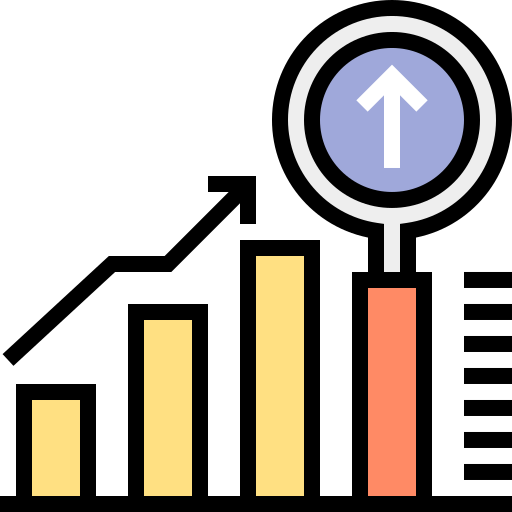}}
\label{subsec:cfa:gold-label-availability}

To \compile an \fmware, the search procedure requires gold labels to determine the quality of each intermediate solution and guide the search toward prompt templates and component parameters that lead to the best solutions~\cite{opsahlong2024optimizinginstructionsdemonstrationsmultistage}.
The software engineering community has developed extensive benchmarking frameworks to evaluate FMs and FM-powered systems across diverse tasks. Prominent benchmarks include HumanEval and MBPP for code generation~\cite{chen2021evaluating,austin2021program}, SWE-bench for realistic software engineering tasks~\cite{jimenez2024swebench}, MMLU for multitask language understanding~\cite{hendrycks2021measuring}, and task-specific datasets for question-answering, summarization, and reasoning. These benchmarks provide standardized evaluation datasets with gold-standard examples that have enabled systematic comparison of different models and approaches.

In the context of FMware compilation, gold labels are associated with the tasks performed by the \fmware component under optimization (see Section~\ref{subsec:cfa:fmware-representation}), being materialized as a set of demonstrations along with their ground truth for evaluating the optimized parameters by the \comp.
Such gold labels can be used at the prompt template level to evaluate a specific template and the parameters of the components associated with that template, or at the \fmware level to jointly evaluate the configuration of all parameters and prompt templates simultaneously.
However, such gold labels are not highly available for the diversity of tasks that software makers can want to perform with an \fmware.
Despite the ability of \acp{fm} to generate such gold labels, research is still required to determine whether (and to which extent) this synthetic generation approach would introduce any negative side effects in the \compilation, particularly given concerns about data contamination where models are evaluated on data similar to their training sets~\cite{deng2024investigating}.

\cfa{The different communities investing in \fmware \ac{randd} should create gold labels with clearly stated assumptions about the data. In particular, such gold labels should be of high quality, with independent data points, and representative of true populations.}

\subsubsection{Quality range estimation  \includegraphics[height=\baselineskip]{figs/crosscut-saw.png}}
\label{subsec:cfa:quality-range-estimation}

Compilation of intents into FMware is not done with the assumption of the deterministic behaviour of the \compiled components.
Hence, intent \comps should include mechanisms to guarantee that the \compiled \fmware executes at pre-defined quality standard levels within a confidence interval.
For example, an \fmware that performs sentiment classification~\cite{sahu2022dataintentclassification} might wish to execute at a specified accuracy range (i.e., the proportion of correctly classified intentions).
Similarly, an \fmware might want to execute under a range of other quality attribute metrics related to safety or fairness, for example.
Such capability can be achieved by sampling these quality attributes during the \compilation process (see Section~\ref{subsec:cfa:gold-label-availability}) and calculating, within a provided confidence level, a range of quality attribute metrics for which the \compiled \fmware operates under many different configurations.
The \compilation process would then search under the constraint of the pre-specified confidence level and quality attribute metrics.
If the \fmware program cannot be \compiled within a specified threshold of quality levels, the \compilation should fail, such that measures can be taken to ``fix'' the program (e.g., by adding an extra component, changing the static parameters of some existing component, or refining the intents expressed with the \fmware programming formalism).

\cfa{\ac{se30} \comps should provide mechanisms to calculate the probability that a program will execute under certain quality thresholds and inform the software maker when such thresholds are not achieved by any of the searched \fmware configurations.}

\subsubsection{Efficiency improvement and cost reduction \includegraphics[height=\baselineskip]{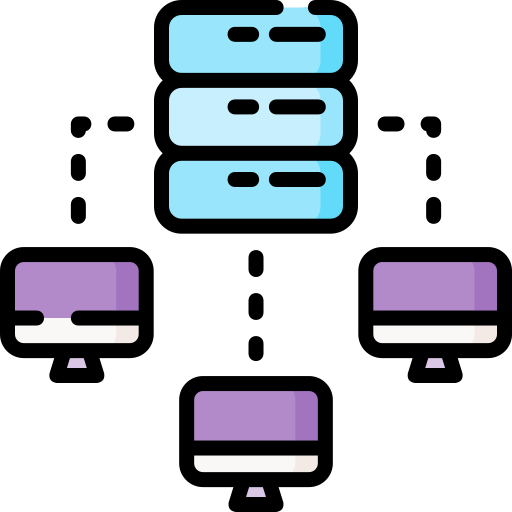}}

\Compiling an \fmware is computationally costly~\cite{taherkhani2024epiccosteffectivesearchbasedprompt}, as it requires multiple evaluations of a performance metric given a candidate configuration of the system, with each evaluation typically requiring one or more inferences with the \ac{fm}.
The high computational (and potentially financial) costs associated with repeated FM calls present a substantial bottleneck for compilation efficiency. Recent advances in LLM inference optimization have demonstrated promising approaches to address these challenges. Prompt compression methods such as LLMLingua achieve up to 20x compression with minimal performance loss by identifying and removing less important tokens~\cite{jiang2023llmlingua}, directly reducing both latency and cost per inference. Additionally, optimizations to the inference process itself, including techniques such as PagedAttention~\cite{kwon2023efficient} and speculative decoding~\cite{leviathan2023fast}, can significantly reduce the per-call overhead of FM interactions.

Hence, the \compilation of an \fmware requires measures to improve the \compilation efficiency and allow the search space to be efficiently and effectively explored.
While these optimization techniques have been developed primarily for improving LLM application performance, they are directly applicable to the compilation context.
General methods such as concurrency, semantic caching~\cite{bang2023gptcache}, and check-pointing have the potential to improve efficiency and reduce costs of \compilation, but research still needs to quantify the improvements of using such methods in the specific context of FMware compilation where compilation traces, candidate solutions, and fitness evaluations present unique caching opportunities.
In addition, specific \compile optimization methods such as operator fusion~\cite{chen2018tvm} and cost modelling~\cite{baghdadi2021dlcostmodel} can be adapted from traditional and DL compilers to \ac{se30} \comps. However, the iterative and exploratory nature of search-based compilation introduces additional considerations: caching must balance between storing expensive computation results and avoiding over-reliance on previously seen solutions, which could limit the diversity of the search process (as discussed in Section~\ref{sec:search-mechanism}).

\cfa{Techniques to reduce the latency and increase the throughput of \compilation should be further developed, such that the parameter search space can be explored more thoroughly for a given computing budget. In particular, research should investigate how semantic caching, prompt compression, and inference optimization techniques can be effectively integrated into search-based compilation workflows while maintaining search diversity. Similarly, techniques to reduce the cost of performing an end-to-end \compilation should be provided and evaluated, with careful attention to the trade-offs between compilation efficiency and solution quality.}

\subsubsection{Reproducible compilations  \includegraphics[height=\baselineskip]{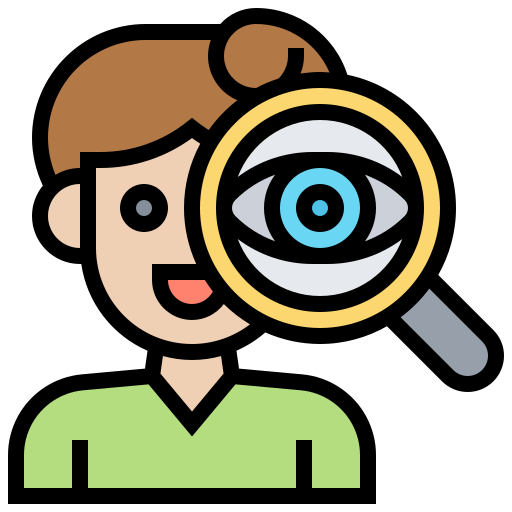}}

Reproducible \compilation is an important feature to allow software makers to build trustworthy \fmware, as it allows the delivered \fmware to be externally verified (e.g. to ensure that the \comp is not tampered with and no adversarial prompts are injected during the \compilation process).
The SE community has developed robust practices around reproducible builds as a set of development practices that create an independently-verifiable path from source to binary code. Reproducible builds enable external parties to verify that distributed binaries match their source code exactly, strengthening software supply chain security and allowing for independent audits~\cite{lamb2022reproducible,decarnedecarnelatmannan2014challenges}. These practices have been successfully adopted by major open-source projects (e.g., Debian, NixOS, Bitcoin Core) and rely on deterministic build systems, stable inputs, pinned dependencies, and well-defined build environments.

Ideally, the \compilation process for \fmware should be reproducible, with the same \fmware program and configuration \compiling to the same output set of parameters and prompt templates~\cite{carnavalet2014verifiablebuilds}.
However, achieving reproducibility in intent compilation presents unique challenges compared to traditional reproducible builds.
Because \acp{fm} are typically used to guide the \compilation process (e.g., to implement prompt mutation heuristics~\cite{fernando2024promptbreeder}), the search process is invariably non-deterministic.
In this case, \ac{fm} decoder parameters (e.g., temperature~\cite{renze2024effectsamplingtemperatureproblem}) can be adjusted to mitigate non-deterministic outputs from the \ac{fm}-generated content, 
making the process more reproducible in practice. However, non-deterministic behavior can still arise from multiple sources (e.g., sampling strategies, runtime environment, or hardware dependencies). This highlights that reproducibility in intent compilation is not only a matter of tuning FM parameters, but rather a broader systems challenge requiring systematic control of all sources of non-determinism. Similar issues are well known in other software engineering domains, such as reproducible builds and machine learning training, where exact replication of outcomes is difficult. Therefore, SE 3.0 compilers should adopt and extend established techniques (e.g., record-and-replay, systematic logging of seeds and runtime states~\cite{decarnedecarnelatmannan2014challenges,lamb2022reproducible}) to mitigate non-deterministic behaviour and make compiled FMware more trustworthy.

Moreover, search algorithms typically require randomness to ensure that various solutions are verified for performance and that the algorithm can escape local optima, posing challenges to the reproducibility of  \compilation.
Therefore, \comps should
balance the need for reproducibility with the need for effective search, potentially through controlled randomness via documented seed values or by ensuring that the compilation trace itself (including all intermediate states and decisions) can be deterministically replayed even if the original search was non-deterministic~\cite{shi2022verifiablebuildslargescalecomm}.
Similar to build~\cite{lamb2022reproduciblebuildssupplechain} and \ac{dl}~\cite{chen2022reproducibledl} training reproducibility, we foresee a window of research challenges and opportunities related to the reproducibility of  \comps.

Finally, the SE 3.0 vision and the compiler infrastructure we propose around it suggest reconsidering how reproducibility is achieved for FMware. Rather than pursuing bit-for-bit identical outputs, which may be unattainable with FMs, \autocog enables quality-level reproducibility in FMware by systematically guaranteeing that recompiled systems meet the same performance thresholds even if exact artifacts differ (see Section~\ref{subsec:cfa:quality-range-estimation}). The compiler provides (a) systematic visibility into quality metrics, making it easy to detect when recompilation fails to meet requirements, (b) automated validation against gold labels to verify functional equivalence, and (c) automated recompilation when models or requirements change, guaranteeing quality restoration without manual prompt adjustments. This shifts reproducibility from ``regenerate exact artifacts'' to ``guarantee quality properties'' which is both more achievable and more valuable for maintaining FMware components over time.

\cfa{Despite the inherent challenges to guarantee determinism, \ac{se30} \comps should implement techniques to make the compilation process reproducible such that the executable \fmware can be verified by external parties. In particular, reproducibility in intent compilation should adopt and extend well-known approaches from reproducible builds (such as deterministic build systems, stable inputs, and systematic documentation) and ML training (such as record-and-replay and systematic logging). Research should investigate how to balance the tension between search diversity and reproducibility, potentially through mechanisms such as deterministic replay of compilation traces or controlled randomness with documented seeds.}

\subsubsection{User-defined optimization objectives  \includegraphics[height=\baselineskip]{figs/search-engine.png}}

During the \compilation of a \fmware, software makers might want to focus on different co-existing optimization objectives, such as the cost of executing the \fmware (e.g., measured in number of tokens), the overall application latency, or the application's performance in a task. 
Similarly to traditional compilers where users can define different optimization preferences, \comps should treat optimization objectives as user-defined parameters, with software makers possibly assigning different weights to each of such objectives.
This feature would require \comps to implement mechanisms to allow users to define the optimization objectives with different levels of flexibility, either with pre-defined objectives directly implemented by the \comp or with a plug-in-based approach where any objective can be defined by the user.   
Such a feature also requires the implementation of multi-objective search algorithms by \comps and the ability to select a single solution from the potential set of competing solutions that dominate each other in different objective dimensions.

\cfa{A flexible and reusable approach to define optimization objectives should be defined by \ac{se30} \comps, which would support the alignment with specific users’ needs and use cases.}



\subsubsection{Interoperability between \comps  \includegraphics[height=\baselineskip]{figs/crosscut-saw.png}}

The number of \ac{se30} \comp offerings will eventually increase, posing emerging interoperability requirements for \fmware programs, so they can be composed using multiple \comp technologies and the same \fmware implementation can be easily ported from one \comp to another, avoiding vendor lock-in.
Cross-language and cross-compiler interoperability has been a longstanding challenge in SE. Traditional approaches have ranged from defining common calling conventions and application binary interfaces~\cite{chisnall2013challenge}, to building multi-language runtimes that share a common intermediate representation~\cite{grimmer2018crosslanguage}. The success of LLVM demonstrates how a well-designed, language-independent IR can enable front-ends for many programming languages to interoperate with back-ends for many target architectures~\cite{LLVM}.

For SE 3.0 compilers, possible approaches to solve the interoperability problem include directly transforming a program representation that is compatible with one \comp to another representation compatible with another \comp~\cite{liu2020enhancinginteropdl}, or proposing a unified \ac{ir} to which all \fmware programs can be converted~\cite{jin2020compilingonnxneuralnetwork} back-and-forth. The latter approach, inspired by LLVM's success in the traditional compiler domain and MLIR's extensible dialect system, appears particularly promising. A standardized FMware IR could define common abstractions for prompts, cognitive architectures, agent coordination, and RAG components while allowing compiler-specific dialects for domain-specific optimizations.

\cfa{It is important that software makers benefit from the strengths of each \ac{se30} \comp offer and can compose programs that are compatible with different \comps. Therefore, efforts should be made to ensure that the different \comps can interoperate, such as establishing a standardized \ac{ir} to which \fmware programs can be transformed.}

\subsubsection{Community-sharing of \compilation traces  \includegraphics[height=\baselineskip]{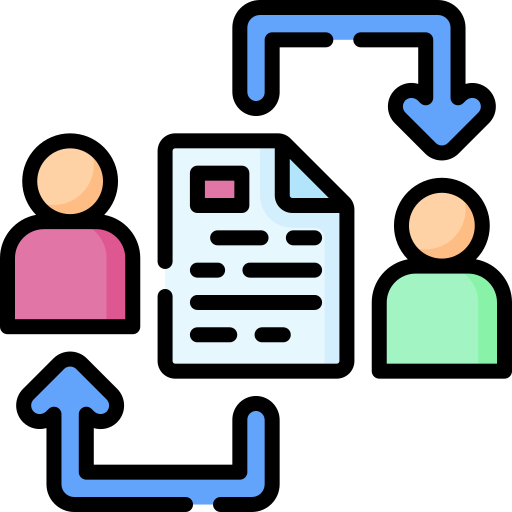}}

The \compilation process generates traces containing valuable information that can be used to continuously improve the \compilation process.
In particular, each \compilation involves different problem definitions (e.g., through the expression of intentions, as proposed in Section~\ref{subsec:cfa:fmware-representation}) and objective functions calculations (e.g., by evaluating task performance using gold labels, as proposed in Section~\ref{subsec:cfa:gold-label-availability}).
If this information is recorded along with the intermediate solutions (i.e., values of the optimized parameters and prompt templates) during the \compilation process, such information can be reused in future compilations that have the same problem definitions and objective function or used to build cost models~\cite{chen2018tvm}, ultimately avoiding re-computation of some of the intermediate solutions and the associated values of the objective function. 
The traces can also include final prompts for a certain task, and they can be used to bootstrap the prompts of other similar \fmware and \acp{fm},
Therefore, \comps will benefit from community sharing of such \compilation traces, as this information can be reused to benefit all users of the same \comp.

\cfa{\Compilation traces contain valuable information that can be used to avoid re-computations during \compilation. As such, \comps should provide resources to allow users to opt-in for sharing of \compilation traces and mechanisms to feedback this information into the \comp to accelerate and improve the \compilation process.}

\subsection{Research Opportunities at the Intersection with Top-Down SE Paradigms}
\label{subsec:research-opportunities-topdown-se}

The challenges outlined in our Calls for Action have parallels with longstanding research in several ``top-down'' SE approaches that transform a higher-level representation of a software system into executable artifacts. While we frame our vision through compilation, these paradigms offer complementary perspectives and proven techniques that can inform SE 3.0 compiler research.

\noindent \textbf{Model-Based Software Engineering.} MBSE emphasizes the systematic generation of implementations from high-level models, using model transformations and well-defined modelling languages to support the entire system lifecycle~\cite{brambilla2017model}. SE 3.0 compilers share this top-down philosophy: both transform abstract specifications (models or intents) into executable implementations. However, the approaches differ in fundamental ways. MBSE typically operates with deterministic, formal models and transformations, while SE 3.0 compilers must navigate the non-deterministic behaviour of FMs and the ambiguities of natural language. Call for Action 1's emphasis on quality constructs for FMware representation parallels MBSE's decades of work on metamodels and modelling language design, research that established principles of abstraction, compositionality, and semantic precision. Similarly, Call for Action 9's focus on interoperability aligns with MBSE's extensive work on model interchange formats and transformation standards. SE 3.0 compilers can adopt proven metamodeling techniques to define FMware IRs and validate FMware programs during compilation.

\noindent \textbf{Self-Adaptive Systems.} Self-adaptive systems employ feedback loops to enable runtime adaptation under changing conditions and uncertainty~\cite{garlan2009software,delemos2013software}. SE 3.0 compilers face analogous challenges at compilation-time: monitoring FMware behaviour through gold labels, analyzing quality metrics, planning optimization strategies, and executing transformations, all guided by accumulated knowledge about effective approaches. Call for Action 3's focus on effective search heuristics aligns with self-adaptive systems' extensive research on adaptation strategies and optimization under uncertainty. Call for Action 6 on efficiency and cost reduction parallels self-adaptive systems' work on resource-aware adaptation and performance optimization. Call for Action 8 on quality assurance shares challenges with verifying self-adaptive systems under non-deterministic conditions, as both domains must ensure correctness despite inherent uncertainty. Research opportunities include adapting MAPE-K~\cite{kephart2003vision} and AWARE~\cite{chekam2025breaking} patterns for compilation-time use, applying self-adaptive verification techniques to FMware testing, and exploring how compilation-time optimization strategies might inform runtime adaptation decisions in deployed FMware systems (e.g., route runtime calls to different FMs based on their performance on the gold labels during compilation).

\noindent \textbf{Software Product Lines (SPLs).} SPL engineering provides systematic methods for managing variability in software families through feature models and automated configuration~\cite{benavides2010automated}. The parallel to SE 3.0 compilers is substantial, as both select from a space of possible configurations to optimize desired properties. In SPL, engineers configure features, while in SE 3.0 compilation, compilers select prompts, models, architectural patterns, and other FMware components. Call for Action 3 on search heuristics directly relates to SPL's extensive literature on feature selection optimization, including genetic algorithms, constraint solving, and multi-objective optimization techniques. Call for Action 8 on quality assurance resonates with SPL testing strategies, particularly sampling techniques for efficiently testing product variants, analogous to testing multiple FMware configurations~\cite{thum2014featureide}. Research opportunities include adapting SPL configuration algorithms to FMware optimization and applying variability testing strategies to assess FMware configurations.

These established SE paradigms provide formalized methods and techniques developed over decades that can inform SE 3.0 compiler research. We encourage researchers from these communities to explore these intersections, bringing their expertise to bear on the emerging challenges of compiling FMware.

\subsection{Bridging intent compilation and SE practices}
\label{subsec:bridging-intent-compilation-with-se-practices}

While this roadmap focuses on challenges related to intent compilation, \autocog integrates with established software engineering practices to support complete development workflows. Below, we present a set of RQs that bridge the compilation infrastructure we envision with broader SE practices to operate effectively in the SE 3.0 era.

\noindent \textbf{Requirements Engineering.} Call for Action 1's quality constructs for intent representation directly address requirements specification in the SE 3.0 context. Intents serve as a form of high-level requirements that must be validated against stakeholder needs, traced through the compilation process to generate FMware artifacts, and evolved as requirements change. Future research should investigate: (i) how intent specifications can be systematically elicited from stakeholders, (ii) how conflicts between multiple intents can be detected and resolved, and (iii) how requirements traceability can be maintained from intents through compilation to deployed FMware.

\noindent \textbf{Software Testing and Quality Assurance.} Call for Action 4's gold labels serve dual purposes: guiding compilation-time optimization and seeding evaluations for deployed FMware. However, compilation-time evaluation represents only a subset of comprehensive testing. Research should investigate: (i) how gold labels used during compilation relate to production test suites, (ii) how test coverage can be systematically expanded beyond the scenarios used during compilation, and (iii) how regression testing can leverage compilation traces to identify when recompilation is necessary.

\noindent \textbf{Continuous Integration and DevOps.} Call for Action 6 (efficiency) and Call for Action 10 (compilation trace sharing) enable integration with CI/CD pipelines. Future work should develop practices for: (i) incremental recompilation triggered by changes to intents or system parameters, (ii) caching strategies that work across development teams and build servers, (iii) quality gates that fail builds when compilation cannot meet specified quality thresholds (Call for Action 5), and (iv) automated deployment pipelines that package compiled FMware artifacts with necessary runtime infrastructure.

\noindent \textbf{Software Maintenance and Evolution.} SE 3.0 systems will evolve as requirements change, new foundation models become available, and feedback accumulates from production deployments. Call for Action 7 (reproducibility) ensures predictable recompilation when systems must be rebuilt, while the search-based approach naturally supports iterative refinement. Research should investigate: (i) how FMware artifacts are maintained over time as dependencies, (ii) how compilation strategies can adapt to model improvements without requiring complete re-optimization, (iii) how technical debt manifests in compiled FMware (e.g., prompts optimized for outdated models), and (iv) how version control and configuration management apply to FMware artifacts.

\noindent \textbf{Collaborative Development.} Modern software development is inherently collaborative, involving multiple developers, designers, domain experts, and stakeholders. Research should investigate how: (i) multiple developers can collaborate on FMware development with shared compilation caches and reusable gold labels, (ii) compilation traces can document design decisions and optimization rationale for knowledge sharing.

\noindent \textbf{Project Management and Process.} SE 3.0 projects require planning, resource allocation, and process management adapted to intent-driven development. Research should investigate: (i) how compilation time and cost can be estimated for project planning, (ii) how compilation-time metrics (convergence rates, quality improvements) can inform project status and risk assessment, and (iii) how agile development practices (sprints, iterative refinement, continuous delivery) adapt to compilation-based workflows.

Finally, in our previous paper~\cite{hassan2024rethinkingse}, we focus on challenges for building trustworthy FMware in practice, highlighting issues such as debugging difficulties, prompt fragility, high operational costs, non-deterministic testing, lack of collaboration support, and siloed tooling, in this paper we discuss a set of calls for action, focusing on the challenges around the research and development of an infrastructure to synthesize FMware programs based on developers’ intents. These challenges include developing quality constructs for representing FMware, effective search heuristics, reproducible and cost-efficient compilation, interoperability between compilers, and sharing of compilation traces.
\section{Conclusion}
\label{sec:conclusions}

\ac{ai}-native systems, especially those powered by \acp{fm}, require continuous adaptation and optimization due to their non-deterministic and evolving nature. Traditional compilers, designed for static environments, cannot handle these real-time adjustments. 
In this context, \autocog’s search-based approach allows it to optimize various aspects of cognitive architectures, such as prompts, \ac{fm} configurations, and system parameters, ensuring that these \ac{ai} systems remain efficient and effective even as they evolve.
By leveraging advanced search-based techniques, \autocog ensures optimal trade-offs between various objectives such as accuracy, latency, and cost, enhancing the efficiency of \ac{ai}-powered systems. 

Our outlined research and development roadmap presents actionable steps for the \ac{se} community to further develop and refine intent compilers, focusing on interoperability, reproducibility, and cost reduction.
Ultimately, intent \comps not only pushes the boundaries of \ac{ai} in \ac{se} but also contributes to a future where software creation is more intuitive, accessible, and aligned with human intent.

\section*{Disclaimer}
Any opinions, findings, conclusions, or recommendations expressed in this material are those of the author(s) and do not reflect the views of Huawei. Also, ChatGPT with GPT-4o was used for copy-editing and table formatting. All experiments, analysis, and results were performed by the authors, who also thoroughly reviewed the final written content for accuracy. This complies with IEEE and ACM policies on AI use in publications.

\balance
\footnotesize
\bibliographystyle{IEEEtranN}
\bibliography{IEEEabrv,references}

\begin{thebibliography}{94}
\providecommand{\natexlab}[1]{#1}
\providecommand{\url}[1]{#1}
\csname url@samestyle\endcsname
\providecommand{\newblock}{\relax}
\providecommand{\bibinfo}[2]{#2}
\providecommand{\BIBentrySTDinterwordspacing}{\spaceskip=0pt\relax}
\providecommand{\BIBentryALTinterwordstretchfactor}{4}
\providecommand{\BIBentryALTinterwordspacing}{\spaceskip=\fontdimen2\font plus
\BIBentryALTinterwordstretchfactor\fontdimen3\font minus \fontdimen4\font\relax}
\providecommand{\BIBforeignlanguage}[2]{{%
\expandafter\ifx\csname l@#1\endcsname\relax
\typeout{** WARNING: IEEEtranN.bst: No hyphenation pattern has been}%
\typeout{** loaded for the language `#1'. Using the pattern for}%
\typeout{** the default language instead.}%
\else
\language=\csname l@#1\endcsname
\fi
#2}}
\providecommand{\BIBdecl}{\relax}
\BIBdecl

\bibitem[Vaithilingam et~al.(2022)Vaithilingam, Zhang, and Glassman]{vaithilingam2022expectation}
P.~Vaithilingam, T.~Zhang, and E.~L. Glassman, ``Expectation vs. experience: Evaluating the usability of code generation tools powered by large language models,'' in \emph{Extended Abstracts of the 2022 CHI Conference on Human Factors in Computing Systems}, ser. CHI EA'22.\hskip 1em plus 0.5em minus 0.4em\relax Association for Computing Machinery, 2022.

\bibitem[Jalil(2023)]{jalil2023transformative}
\BIBentryALTinterwordspacing
S.~Jalil, ``The transformative influence of large language models on software development,'' 2023. [Online]. Available: \url{https://arxiv.org/abs/2311.16429}
\BIBentrySTDinterwordspacing

\bibitem[Zhang et~al.(2023)Zhang, Liang, Zhou, Ahmad, and Waseem]{zhang2023practices}
B.~Zhang, P.~Liang, X.~Zhou, A.~Ahmad, and M.~Waseem, ``Practices and challenges of using github copilot: An empirical study,'' in \emph{Proceedings of the 35th International Conference on Software Engineering and Knowledge Engineering}, ser. SEKE2023, vol. 2023.\hskip 1em plus 0.5em minus 0.4em\relax KSI Research Inc., 2023, p. 124–129.

\bibitem[Zhou et~al.(2025)Zhou, Liang, Zhang, Li, Ahmad, Shahin, and Waseem]{zhou2025exploring}
X.~Zhou, P.~Liang, B.~Zhang, Z.~Li, A.~Ahmad, M.~Shahin, and M.~Waseem, ``Exploring the problems, their causes and solutions of ai pair programming: A study on github and stack overflow,'' \emph{J. Syst. Softw.}, vol. 219, no.~C, 2025.

\bibitem[Jimenez et~al.(2024)Jimenez, Yang, Wettig, Yao, Pei, Press, and Narasimhan]{jimenez2024swebench}
\BIBentryALTinterwordspacing
C.~E. Jimenez, J.~Yang, A.~Wettig, S.~Yao, K.~Pei, O.~Press, and K.~Narasimhan, ``Swe-bench: Can language models resolve real-world github issues?'' 2024. [Online]. Available: \url{https://arxiv.org/abs/2310.06770}
\BIBentrySTDinterwordspacing

\bibitem[Yuan et~al.(2024)Yuan, Liu, Ding, Wang, Chen, Peng, and Lou]{yuan2024evaluating}
Z.~Yuan, M.~Liu, S.~Ding, K.~Wang, Y.~Chen, X.~Peng, and Y.~Lou, ``Evaluating and improving chatgpt for unit test generation,'' \emph{Proc. ACM Softw. Eng.}, vol.~1, no. FSE, Jul. 2024.

\bibitem[Guo et~al.(2024{\natexlab{a}})Guo, Cao, Xie, Liu, Li, Chen, and Peng]{guo2024exploring}
Q.~Guo, J.~Cao, X.~Xie, S.~Liu, X.~Li, B.~Chen, and X.~Peng, ``Exploring the potential of chatgpt in automated code refinement: An empirical study,'' in \emph{Proceedings of the IEEE/ACM 46th International Conference on Software Engineering}, ser. ICSE '24.\hskip 1em plus 0.5em minus 0.4em\relax New York, NY, USA: Association for Computing Machinery, 2024.

\bibitem[Hassan et~al.(2024{\natexlab{a}})Hassan, Oliva, Lin, Chen, Ming, and Jiang]{hassan2024ainativesoftwareengineeringse}
\BIBentryALTinterwordspacing
A.~E. Hassan, G.~A. Oliva, D.~Lin, B.~Chen, Z.~Ming, and Jiang, ``Towards ai-native software engineering (se 3.0): A vision and a challenge roadmap,'' 2024. [Online]. Available: \url{https://arxiv.org/abs/2410.06107}
\BIBentrySTDinterwordspacing

\bibitem[Li et~al.(2025)Li, Zhang, and Hassan]{li2025riseaiteammatessoftware}
\BIBentryALTinterwordspacing
H.~Li, H.~Zhang, and A.~E. Hassan, ``The rise of ai teammates in software engineering (se) 3.0: How autonomous coding agents are reshaping software engineering,'' 2025. [Online]. Available: \url{https://arxiv.org/abs/2507.15003}
\BIBentrySTDinterwordspacing

\bibitem[Sclar et~al.(2024)Sclar, Choi, Tsvetkov, and Suhr]{sclar2024quantifying}
\BIBentryALTinterwordspacing
M.~Sclar, Y.~Choi, Y.~Tsvetkov, and A.~Suhr, ``Quantifying language models' sensitivity to spurious features in prompt design or: How i learned to start worrying about prompt formatting,'' in \emph{The Twelfth International Conference on Learning Representations}, 2024. [Online]. Available: \url{https://openreview.net/forum?id=RIu5lyNXjT}
\BIBentrySTDinterwordspacing

\bibitem[Cao et~al.(2024)Cao, Cai, Zhang, Zou, and Lam]{cao2024worstpromptperformancelarge}
\BIBentryALTinterwordspacing
B.~Cao, D.~Cai, Z.~Zhang, Y.~Zou, and W.~Lam, ``On the worst prompt performance of large language models,'' 2024. [Online]. Available: \url{https://arxiv.org/abs/2406.10248}
\BIBentrySTDinterwordspacing

\bibitem[Lu et~al.(2022)Lu, Bartolo, Moore, Riedel, and Stenetorp]{lu2022fantasticallyorderedprompts}
\BIBentryALTinterwordspacing
Y.~Lu, M.~Bartolo, A.~Moore, S.~Riedel, and P.~Stenetorp, ``Fantastically ordered prompts and where to find them: Overcoming few-shot prompt order sensitivity,'' in \emph{Proceedings of the 60th Annual Meeting of the Association for Computational Linguistics (Volume 1: Long Papers)}, S.~Muresan, P.~Nakov, and A.~Villavicencio, Eds.\hskip 1em plus 0.5em minus 0.4em\relax Dublin, Ireland: Association for Computational Linguistics, May 2022, pp. 8086--8098. [Online]. Available: \url{https://aclanthology.org/2022.acl-long.556}
\BIBentrySTDinterwordspacing

\bibitem[Hu et~al.(2024{\natexlab{a}})Hu, Lu, and Clune]{hu2024automateddesignagenticsystems}
\BIBentryALTinterwordspacing
S.~Hu, C.~Lu, and J.~Clune, ``Automated design of agentic systems,'' 2024. [Online]. Available: \url{https://arxiv.org/abs/2408.08435}
\BIBentrySTDinterwordspacing

\bibitem[Aho et~al.(1986)Aho, Sethi, and Ullman]{aho1986compilers}
A.~V. Aho, R.~Sethi, and J.~D. Ullman, \emph{Compilers: principles, techniques, and tools}.\hskip 1em plus 0.5em minus 0.4em\relax USA: Addison-Wesley Longman Publishing Co., Inc., 1986.

\bibitem[{Free Software Foundation}(2024)]{gcc2024}
\BIBentryALTinterwordspacing
{Free Software Foundation}, \emph{{GCC} Online Documentation, Version 14.2.0}, 2024, accessed: 2024-10-10. [Online]. Available: \url{https://gcc.gnu.org/onlinedocs/gcc-14.2.0/gcc/}
\BIBentrySTDinterwordspacing

\bibitem[Lattner and Adve(2004)]{LLVM}
C.~Lattner and V.~Adve, ``Llvm: A compilation framework for lifelong program analysis \& transformation,'' in \emph{Proceedings of the International Symposium on Code Generation and Optimization: Feedback-Directed and Runtime Optimization}, 2004, p.~75.

\bibitem[Chen et~al.(2018)Chen, Moreau, Jiang, Zheng, Yan, Cowan, Shen, Wang, Hu, Ceze, Guestrin, and Krishnamurthy]{chen2018tvm}
T.~Chen, T.~Moreau, Z.~Jiang, L.~Zheng, E.~Yan, M.~Cowan, H.~Shen, L.~Wang, Y.~Hu, L.~Ceze, C.~Guestrin, and A.~Krishnamurthy, ``Tvm: an automated end-to-end optimizing compiler for deep learning,'' in \emph{Proceedings of the 13th USENIX Conference on Operating Systems Design and Implementation}, ser. OSDI'18.\hskip 1em plus 0.5em minus 0.4em\relax USA: USENIX Association, 2018, p. 579–594.

\bibitem[{OpenXLA Community}(2024)]{openxla2024}
\BIBentryALTinterwordspacing
{OpenXLA Community}, ``Openxla: Accelerated machine learning compilation,'' 2024, accessed: 2024-10-10. [Online]. Available: \url{https://openxla.org/}
\BIBentrySTDinterwordspacing

\bibitem[Rotem et~al.(2019)Rotem, Fix, Abdulrasool, Catron, Deng, Dzhabarov, Gibson, Hegeman, Lele, Levenstein, Montgomery, Maher, Nadathur, Olesen, Park, Rakhov, Smelyanskiy, and Wang]{rotem2019glowgraphloweringcompiler}
\BIBentryALTinterwordspacing
N.~Rotem, J.~Fix, S.~Abdulrasool, G.~Catron, S.~Deng, R.~Dzhabarov, N.~Gibson, J.~Hegeman, M.~Lele, R.~Levenstein, J.~Montgomery, B.~Maher, S.~Nadathur, J.~Olesen, J.~Park, A.~Rakhov, M.~Smelyanskiy, and M.~Wang, ``Glow: Graph lowering compiler techniques for neural networks,'' 2019. [Online]. Available: \url{https://arxiv.org/abs/1805.00907}
\BIBentrySTDinterwordspacing

\bibitem[Dean et~al.(2012)Dean, Corrado, Monga, Chen, Devin, Le, Mao, Ranzato, Senior, Tucker, Yang, and Ng]{dean2012largescale}
J.~Dean, G.~S. Corrado, R.~Monga, K.~Chen, M.~Devin, Q.~V. Le, M.~Z. Mao, M.~Ranzato, A.~Senior, P.~Tucker, K.~Yang, and A.~Y. Ng, ``Large scale distributed deep networks,'' in \emph{Proceedings of the 25th International Conference on Neural Information Processing Systems - Volume 1}, ser. NIPS'12.\hskip 1em plus 0.5em minus 0.4em\relax Red Hook, NY, USA: Curran Associates Inc., 2012, p. 1223–1231.

\bibitem[Zhou et~al.(2023)Zhou, Muresanu, Han, Paster, Pitis, Chan, and Ba]{zhou2023large}
\BIBentryALTinterwordspacing
Y.~Zhou, A.~I. Muresanu, Z.~Han, K.~Paster, S.~Pitis, H.~Chan, and J.~Ba, ``Large language models are human-level prompt engineers,'' in \emph{The Eleventh International Conference on Learning Representations}, 2023. [Online]. Available: \url{https://openreview.net/forum?id=92gvk82DE-}
\BIBentrySTDinterwordspacing

\bibitem[Fernando et~al.(2024)Fernando, Banarse, Michalewski, Osindero, and Rockt{\"a}schel]{fernando2024promptbreeder}
\BIBentryALTinterwordspacing
C.~Fernando, D.~S. Banarse, H.~Michalewski, S.~Osindero, and T.~Rockt{\"a}schel, ``Promptbreeder: Self-referential self-improvement via prompt evolution,'' in \emph{Forty-first International Conference on Machine Learning}, 2024. [Online]. Available: \url{https://openreview.net/forum?id=9ZxnPZGmPU}
\BIBentrySTDinterwordspacing

\bibitem[Guo et~al.(2024{\natexlab{b}})Guo, Wang, Guo, Li, Song, Tan, Liu, Bian, and Yang]{guo2024connectingevoprompt}
\BIBentryALTinterwordspacing
Q.~Guo, R.~Wang, J.~Guo, B.~Li, K.~Song, X.~Tan, G.~Liu, J.~Bian, and Y.~Yang, ``Connecting large language models with evolutionary algorithms yields powerful prompt optimizers,'' in \emph{The Twelfth International Conference on Learning Representations}, 2024. [Online]. Available: \url{https://openreview.net/forum?id=ZG3RaNIsO8}
\BIBentrySTDinterwordspacing

\bibitem[Pryzant et~al.(2023)Pryzant, Iter, Li, Lee, Zhu, and Zeng]{pryzant2023automaticpromptoptimizationwithgradientdescent}
\BIBentryALTinterwordspacing
R.~Pryzant, D.~Iter, J.~Li, Y.~Lee, C.~Zhu, and M.~Zeng, ``Automatic prompt optimization with {``}gradient descent{''} and beam search,'' in \emph{Proceedings of the 2023 Conference on Empirical Methods in Natural Language Processing}, H.~Bouamor, J.~Pino, and K.~Bali, Eds.\hskip 1em plus 0.5em minus 0.4em\relax Singapore: Association for Computational Linguistics, Dec. 2023, pp. 7957--7968. [Online]. Available: \url{https://aclanthology.org/2023.emnlp-main.494}
\BIBentrySTDinterwordspacing

\bibitem[Schnabel and Neville(2024)]{schnabel2024prompts}
\BIBentryALTinterwordspacing
T.~Schnabel and J.~Neville, ``Prompts as programs: A structure-aware approach to efficient compile-time prompt optimization,'' April 2024. [Online]. Available: \url{https://www.microsoft.com/en-us/research/publication/prompts-as-programs-a-structure-aware-approach-to-efficient-compile-time-prompt-optimization/}
\BIBentrySTDinterwordspacing

\bibitem[Hassan et~al.(2024{\natexlab{b}})Hassan, Lin, Rajbahadur, Gallaba, Cogo, Chen, Zhang, Thangarajah, Oliva, Lin, Abdullah, and Jiang]{hassan2024rethinkingse}
\BIBentryALTinterwordspacing
A.~E. Hassan, D.~Lin, G.~K. Rajbahadur, K.~Gallaba, F.~R. Cogo, B.~Chen, H.~Zhang, K.~Thangarajah, G.~Oliva, J.~J. Lin, W.~M. Abdullah, and Z.~M.~J. Jiang, ``Rethinking software engineering in the era of foundation models: A curated catalogue of challenges in the development of trustworthy fmware,'' in \emph{Companion Proceedings of the 32nd ACM International Conference on the Foundations of Software Engineering}, ser. FSE 2024.\hskip 1em plus 0.5em minus 0.4em\relax New York, NY, USA: Association for Computing Machinery, 2024, p. 294–305. [Online]. Available: \url{https://doi.org/10.1145/3663529.3663849}
\BIBentrySTDinterwordspacing

\bibitem[Park et~al.(2023)Park, O'Brien, Cai, Morris, Liang, and Bernstein]{park2023generativeagents}
\BIBentryALTinterwordspacing
J.~S. Park, J.~O'Brien, C.~J. Cai, M.~R. Morris, P.~Liang, and M.~S. Bernstein, ``Generative agents: Interactive simulacra of human behavior,'' in \emph{Proceedings of the 36th Annual ACM Symposium on User Interface Software and Technology}, ser. UIST '23.\hskip 1em plus 0.5em minus 0.4em\relax New York, NY, USA: Association for Computing Machinery, 2023. [Online]. Available: \url{https://doi.org/10.1145/3586183.3606763}
\BIBentrySTDinterwordspacing

\bibitem[Yao et~al.(2023)Yao, Zhao, Yu, Du, Shafran, Narasimhan, and Cao]{yao2023react}
\BIBentryALTinterwordspacing
S.~Yao, J.~Zhao, D.~Yu, N.~Du, I.~Shafran, K.~R. Narasimhan, and Y.~Cao, ``React: Synergizing reasoning and acting in language models,'' in \emph{The Eleventh International Conference on Learning Representations}, 2023. [Online]. Available: \url{https://openreview.net/forum?id=WE_vluYUL-X}
\BIBentrySTDinterwordspacing

\bibitem[Microsoft(2023)]{mspromptflow}
Microsoft, ``Prompt flow documentation,'' \url{https://microsoft.github.io/promptflow}, 2023, accessed 02-06-2024.

\bibitem[Wu et~al.(2023{\natexlab{a}})Wu, Bansal, Zhang, Wu, Li, Zhu, Jiang, Zhang, Zhang, Liu, Awadallah, White, Burger, and Wang]{wu2023autogenenablingnextgenllm}
\BIBentryALTinterwordspacing
Q.~Wu, G.~Bansal, J.~Zhang, Y.~Wu, B.~Li, E.~Zhu, L.~Jiang, X.~Zhang, S.~Zhang, J.~Liu, A.~H. Awadallah, R.~W. White, D.~Burger, and C.~Wang, ``Autogen: Enabling next-gen llm applications via multi-agent conversation,'' 2023. [Online]. Available: \url{https://arxiv.org/abs/2308.08155}
\BIBentrySTDinterwordspacing

\bibitem[Sculley et~al.(2015)Sculley, Holt, Golovin, Davydov, Phillips, Ebner, Chaudhary, Young, Crespo, and Dennison]{Sculley15}
D.~Sculley, G.~Holt, D.~Golovin, E.~Davydov, T.~Phillips, D.~Ebner, V.~Chaudhary, M.~Young, J.-F. Crespo, and D.~Dennison, ``Hidden technical debt in machine learning systems,'' in \emph{Proceedings of the 28th International Conference on Neural Information Processing Systems - Volume 2}, ser. NIPS'15.\hskip 1em plus 0.5em minus 0.4em\relax Cambridge, MA, USA: MIT Press, 2015, p. 2503–2511.

\bibitem[ope(2024)]{opea2024}
``Open platform for enterprise ai (opea),'' \url{https://opea.dev/}, 2024, accessed: 2024-10-11.

\bibitem[{Amazon Web Services (AWS)}()]{DataFlywheel}
\BIBentryALTinterwordspacing
{Amazon Web Services (AWS)}, ``The data flywheel,'' accessed: 2024-09-29. [Online]. Available: \url{https://pages.awscloud.com/EMEA-Data-Flywheel.html}
\BIBentrySTDinterwordspacing

\bibitem[Khattab et~al.(2023)Khattab, Singhvi, Maheshwari, Zhang, Santhanam, Vardhamanan, Haq, Sharma, Joshi, Moazam, Miller, Zaharia, and Potts]{khattab2023dspy}
\BIBentryALTinterwordspacing
O.~Khattab, A.~Singhvi, P.~Maheshwari, Z.~Zhang, K.~Santhanam, S.~Vardhamanan, S.~Haq, A.~Sharma, T.~T. Joshi, H.~Moazam, H.~Miller, M.~Zaharia, and C.~Potts, ``Dspy: Compiling declarative language model calls into self-improving pipelines,'' 2023. [Online]. Available: \url{https://arxiv.org/abs/2310.03714}
\BIBentrySTDinterwordspacing

\bibitem[Yuksekgonul et~al.(2025)Yuksekgonul, Bianchi, Boen, Liu, Huang, Guestrin, and Zou]{yuksekgonul2024textgradautomaticdifferentiationtext}
\BIBentryALTinterwordspacing
M.~Yuksekgonul, F.~Bianchi, J.~Boen, S.~Liu, Z.~Huang, C.~Guestrin, and J.~Zou, ``Optimizing generative ai by backpropagating language model feedback,'' \emph{Nature}, vol. 639, pp. 609--616, 2025. [Online]. Available: \url{https://doi.org/10.1038/s41586-025-08661-4}
\BIBentrySTDinterwordspacing

\bibitem[Schmidt et~al.(2024)Schmidt, Spencer-Smith, Fu, and White]{schmidt2024promptcatalog}
\BIBentryALTinterwordspacing
D.~C. Schmidt, J.~Spencer-Smith, Q.~Fu, and J.~White, ``Towards a catalog of prompt patterns to enhance the discipline of prompt engineering,'' \emph{Ada Lett.}, vol.~43, no.~2, p. 43–51, jun 2024. [Online]. Available: \url{https://doi.org/10.1145/3672359.3672364}
\BIBentrySTDinterwordspacing

\bibitem[Ong et~al.(2024)Ong, Almahairi, Wu, Chiang, Wu, Gonzalez, Kadous, and Stoica]{RouteLLM}
\BIBentryALTinterwordspacing
I.~Ong, A.~Almahairi, V.~Wu, W.-L. Chiang, T.~Wu, J.~E. Gonzalez, M.~W. Kadous, and I.~Stoica, ``Routellm: Learning to route llms with preference data,'' 2024. [Online]. Available: \url{https://arxiv.org/abs/2406.18665}
\BIBentrySTDinterwordspacing

\bibitem[Hu et~al.(2024{\natexlab{b}})Hu, Bieker, Li, Jiang, Keigwin, Ranganath, Keutzer, and Upadhyay]{RouterBench}
\BIBentryALTinterwordspacing
Q.~J. Hu, J.~Bieker, X.~Li, N.~Jiang, B.~Keigwin, G.~Ranganath, K.~Keutzer, and S.~K. Upadhyay, ``Routerbench: A benchmark for multi-llm routing system,'' 2024. [Online]. Available: \url{https://arxiv.org/abs/2403.12031}
\BIBentrySTDinterwordspacing

\bibitem[Eghbali and Pradel(2023)]{aryaz2024crystalbleu}
\BIBentryALTinterwordspacing
A.~Eghbali and M.~Pradel, ``Crystalbleu: Precisely and efficiently measuring the similarity of code,'' in \emph{Proceedings of the 37th IEEE/ACM International Conference on Automated Software Engineering}, ser. ASE '22.\hskip 1em plus 0.5em minus 0.4em\relax New York, NY, USA: Association for Computing Machinery, 2023. [Online]. Available: \url{https://doi.org/10.1145/3551349.3556903}
\BIBentrySTDinterwordspacing

\bibitem[Pan et~al.(2024)Pan, Ibrahimzada, Krishna, Sankar, Wassi, Merler, Sobolev, Pavuluri, Sinha, and Jabbarvand]{pan2024lostintranslation}
\BIBentryALTinterwordspacing
R.~Pan, A.~R. Ibrahimzada, R.~Krishna, D.~Sankar, L.~P. Wassi, M.~Merler, B.~Sobolev, R.~Pavuluri, S.~Sinha, and R.~Jabbarvand, ``Lost in translation: A study of bugs introduced by large language models while translating code,'' in \emph{Proceedings of the IEEE/ACM 46th International Conference on Software Engineering}, ser. ICSE '24.\hskip 1em plus 0.5em minus 0.4em\relax New York, NY, USA: Association for Computing Machinery, 2024. [Online]. Available: \url{https://doi.org/10.1145/3597503.3639226}
\BIBentrySTDinterwordspacing

\bibitem[Bergstra and Bengio(2012)]{bergstra2012randomsearch}
J.~Bergstra and Y.~Bengio, ``Random search for hyper-parameter optimization,'' \emph{J. Mach. Learn. Res.}, vol.~13, no. null, p. 281–305, Feb. 2012.

\bibitem[Wu et~al.(2023{\natexlab{b}})Wu, Yu, Wang, Song, Zhang, Zhao, Lu, Li, and Henao]{wu2023infopromptsoftprompt}
\BIBentryALTinterwordspacing
J.~Wu, T.~Yu, R.~Wang, Z.~Song, R.~Zhang, H.~Zhao, C.~Lu, S.~Li, and R.~Henao, ``Infoprompt: Information-theoretic soft prompt tuning for natural language understanding,'' in \emph{Advances in Neural Information Processing Systems}, A.~Oh, T.~Naumann, A.~Globerson, K.~Saenko, M.~Hardt, and S.~Levine, Eds., vol.~36.\hskip 1em plus 0.5em minus 0.4em\relax Curran Associates, Inc., 2023, pp. 61\,060--61\,084. [Online]. Available: \url{https://proceedings.neurips.cc/paper_files/paper/2023/file/c01c0da4fe2ef2df9863f55261e2e924-Paper-Conference.pdf}
\BIBentrySTDinterwordspacing

\bibitem[Langchain-Ai()]{Langchain-Ai}
\BIBentryALTinterwordspacing
Langchain-Ai, ``Langchain-ai/langchain: Build context-aware reasoning applications.'' [Online]. Available: \url{https://github.com/langchain-ai/langchain}
\BIBentrySTDinterwordspacing

\bibitem[Deb et~al.(2002)Deb, Pratap, Agarwal, and Meyarivan]{deb2002nsgaii}
K.~Deb, A.~Pratap, S.~Agarwal, and T.~Meyarivan, ``A fast and elitist multiobjective genetic algorithm: Nsga-ii,'' \emph{IEEE Transactions on Evolutionary Computation}, vol.~6, no.~2, pp. 182--197, 2002.

\bibitem[Blank and Deb(2020)]{blank2020pymoo}
J.~Blank and K.~Deb, ``Pymoo: Multi-objective optimization in python,'' \emph{IEEE Access}, vol.~8, pp. 89\,497--89\,509, 2020.

\bibitem[Eiben and Smith(2003)]{eiben2003introtoevolcomputing}
A.~Eiben and J.~Smith, \emph{Introduction to Evolutionary Computing}, ser. Natural Computing.\hskip 1em plus 0.5em minus 0.4em\relax Springer-Verlag, Berlin, 2003.

\bibitem[Shinn et~al.(2023)Shinn, Cassano, Gopinath, Narasimhan, and Yao]{shinn2023reflexion}
\BIBentryALTinterwordspacing
N.~Shinn, F.~Cassano, A.~Gopinath, K.~Narasimhan, and S.~Yao, ``Reflexion: language agents with verbal reinforcement learning,'' in \emph{Advances in Neural Information Processing Systems}, A.~Oh, T.~Naumann, A.~Globerson, K.~Saenko, M.~Hardt, and S.~Levine, Eds., vol.~36.\hskip 1em plus 0.5em minus 0.4em\relax Curran Associates, Inc., 2023, pp. 8634--8652. [Online]. Available: \url{https://proceedings.neurips.cc/paper_files/paper/2023/file/1b44b878bb782e6954cd888628510e90-Paper-Conference.pdf}
\BIBentrySTDinterwordspacing

\bibitem[Jebari et~al.(2013)Jebari, Madiafi, et~al.]{jebari2013selection}
K.~Jebari, M.~Madiafi \emph{et~al.}, ``Selection methods for genetic algorithms,'' \emph{International Journal of Emerging Sciences}, vol.~3, no.~4, pp. 333--344, 2013.

\bibitem[Taherkhani et~al.(2024)Taherkhani, Sepindband, Pham, Wang, and Hemmati]{taherkhani2024epiccosteffectivesearchbasedprompt}
\BIBentryALTinterwordspacing
H.~Taherkhani, M.~Sepindband, H.~V. Pham, S.~Wang, and H.~Hemmati, ``Epic: Cost-effective search-based prompt engineering of llms for code generation,'' 2024. [Online]. Available: \url{https://arxiv.org/abs/2408.11198}
\BIBentrySTDinterwordspacing

\bibitem[Hao~Li(2024)]{li2024seandfmindustryblog}
A.~E.~H. Hao~Li, Cor-Paul~Bezemer, ``Software engineering and foundation models: Insights from industry blogs using a jury of foundation models,'' 2024.

\bibitem[Levi et~al.(2024)Levi, Brosh, and Friedmann]{levi2024intentbasedpromptcalibrationenhancing}
\BIBentryALTinterwordspacing
E.~Levi, E.~Brosh, and M.~Friedmann, ``Intent-based prompt calibration: Enhancing prompt optimization with synthetic boundary cases,'' 2024. [Online]. Available: \url{https://arxiv.org/abs/2402.03099}
\BIBentrySTDinterwordspacing

\bibitem[Wang et~al.(2023)Wang, Wei, Schuurmans, Le, Chi, Narang, Chowdhery, and Zhou]{wang2023selfconsistency}
\BIBentryALTinterwordspacing
X.~Wang, J.~Wei, D.~Schuurmans, Q.~V. Le, E.~H. Chi, S.~Narang, A.~Chowdhery, and D.~Zhou, ``Self-consistency improves chain of thought reasoning in language models,'' in \emph{The Eleventh International Conference on Learning Representations}, 2023. [Online]. Available: \url{https://openreview.net/forum?id=1PL1NIMMrw}
\BIBentrySTDinterwordspacing

\bibitem[Anwar et~al.(2024)Anwar, Saparov, Rando, Paleka, Turpin, Hase, Lubana, Jenner, Casper, Sourbut, Edelman, Zhang, G{\"u}nther, Korinek, Hernandez-Orallo, Hammond, Bigelow, Pan, Langosco, Korbak, Zhang, Zhong, hEigeartaigh, Recchia, Corsi, Chan, Anderljung, Edwards, Petrov, de~Witt, Motwani, Bengio, Chen, Torr, Albanie, Maharaj, Foerster, Tram{\`e}r, He, Kasirzadeh, Choi, and Krueger]{anwar2024foundational}
\BIBentryALTinterwordspacing
U.~Anwar, A.~Saparov, J.~Rando, D.~Paleka, M.~Turpin, P.~Hase, E.~S. Lubana, E.~Jenner, S.~Casper, O.~Sourbut, B.~L. Edelman, Z.~Zhang, M.~G{\"u}nther, A.~Korinek, J.~Hernandez-Orallo, L.~Hammond, E.~J. Bigelow, A.~Pan, L.~Langosco, T.~Korbak, H.~C. Zhang, R.~Zhong, S.~O. hEigeartaigh, G.~Recchia, G.~Corsi, A.~Chan, M.~Anderljung, L.~Edwards, A.~Petrov, C.~S. de~Witt, S.~R. Motwani, Y.~Bengio, D.~Chen, P.~Torr, S.~Albanie, T.~Maharaj, J.~N. Foerster, F.~Tram{\`e}r, H.~He, A.~Kasirzadeh, Y.~Choi, and D.~Krueger, ``Foundational challenges in assuring alignment and safety of large language models,'' \emph{Transactions on Machine Learning Research}, 2024. [Online]. Available: \url{https://openreview.net/forum?id=oVTkOs8Pka}
\BIBentrySTDinterwordspacing

\bibitem[Lewis et~al.(2020)Lewis, Perez, Piktus, Petroni, Karpukhin, Goyal, K\"{u}ttler, Lewis, Yih, Rockt\"{a}schel, Riedel, and Kiela]{lewis2021rag}
P.~Lewis, E.~Perez, A.~Piktus, F.~Petroni, V.~Karpukhin, N.~Goyal, H.~K\"{u}ttler, M.~Lewis, W.-t. Yih, T.~Rockt\"{a}schel, S.~Riedel, and D.~Kiela, ``Retrieval-augmented generation for knowledge-intensive nlp tasks,'' in \emph{Proceedings of the 34th International Conference on Neural Information Processing Systems}, ser. NIPS '20.\hskip 1em plus 0.5em minus 0.4em\relax Red Hook, NY, USA: Curran Associates Inc., 2020.

\bibitem[Romero et~al.(2023)Romero, Zimmerman, Steinfeld, and Tomasic]{romero2023synergistic}
O.~J. Romero, J.~Zimmerman, A.~Steinfeld, and A.~Tomasic, ``Synergistic integration of large language models and cognitive architectures for robust ai: An exploratory analysis,'' in \emph{Proceedings of the AAAI Symposium Series}, vol.~2, no.~1.\hskip 1em plus 0.5em minus 0.4em\relax AAAI Press, 2023, pp. 396--405.

\bibitem[Sumers et~al.(2024)Sumers, Yao, Narasimhan, and Griffiths]{sumers2024cognitive}
\BIBentryALTinterwordspacing
T.~Sumers, S.~Yao, K.~Narasimhan, and T.~Griffiths, ``Cognitive architectures for language agents,'' \emph{Transactions on Machine Learning Research}, 2024, survey Certification. [Online]. Available: \url{https://openreview.net/forum?id=1i6ZCvflQJ}
\BIBentrySTDinterwordspacing

\bibitem[Josifoski et~al.(2024)Josifoski, Klein, Peyrard, Baldwin, Li, Geng, Schnitzler, Yao, Wei, Paul, and West]{josifoski2024flowsbuildingblocksreasoning}
\BIBentryALTinterwordspacing
M.~Josifoski, L.~Klein, M.~Peyrard, N.~Baldwin, Y.~Li, S.~Geng, J.~P. Schnitzler, Y.~Yao, J.~Wei, D.~Paul, and R.~West, ``Flows: Building blocks of reasoning and collaborating ai,'' 2024. [Online]. Available: \url{https://arxiv.org/abs/2308.01285}
\BIBentrySTDinterwordspacing

\bibitem[Harman et~al.(2012)Harman, Mansouri, and Zhang]{harman2012search}
\BIBentryALTinterwordspacing
M.~Harman, A.~Mansouri, and Y.~Zhang, ``Search based software engineering: Trends, techniques and applications,'' \emph{ACM Computing Surveys}, vol.~45, no.~1, p. Article 11, 2012. [Online]. Available: \url{https://dl.acm.org/doi/10.1145/2379776.2379787}
\BIBentrySTDinterwordspacing

\bibitem[Afzal et~al.(2009)Afzal, Torkar, and Feldt]{afzal2009asystematic}
W.~Afzal, R.~Torkar, and R.~Feldt, ``A systematic review of search-based testing for non-functional system properties,'' \emph{Information and Software Technology}, vol.~51, no.~6, pp. 957--976, 2009.

\bibitem[Mariani and Vergilio(2017)]{mariani2017asystematic}
T.~Mariani and S.~R. Vergilio, ``A systematic review on search-based refactoring,'' \emph{Information and Software Technology}, vol.~83, pp. 14--34, 2017.

\bibitem[Gulwani et~al.(2017)Gulwani, Polozov, and Singh]{gulwani2017program}
S.~Gulwani, O.~Polozov, and R.~Singh, ``Program synthesis,'' \emph{Foundations and Trends in Programming Languages}, vol.~4, no. 1-2, pp. 1--119, 2017.

\bibitem[Kant(2018)]{kant2018recent}
\BIBentryALTinterwordspacing
N.~Kant, ``Recent advances in neural program synthesis,'' arXiv preprint arXiv:1802.02353, 2018. [Online]. Available: \url{https://arxiv.org/abs/1802.02353}
\BIBentrySTDinterwordspacing

\bibitem[Errica et~al.(2024)Errica, Siracusano, Sanvito, and Bifulco]{errica2024didiwrongquantifying}
\BIBentryALTinterwordspacing
F.~Errica, G.~Siracusano, D.~Sanvito, and R.~Bifulco, ``What did i do wrong? quantifying llms' sensitivity and consistency to prompt engineering,'' 2024. [Online]. Available: \url{https://arxiv.org/abs/2406.12334}
\BIBentrySTDinterwordspacing

\bibitem[Hanna et~al.(2025)Hanna, Blot, and Petke]{hanna2025reinforcement}
\BIBentryALTinterwordspacing
C.~Hanna, A.~Blot, and J.~Petke, ``Reinforcement learning for mutation operator selection in automated program repair,'' \emph{Automated Software Engineering}, vol.~32, 2025. [Online]. Available: \url{https://link.springer.com/article/10.1007/s10515-025-00501-z}
\BIBentrySTDinterwordspacing

\bibitem[Sahin et~al.(2026)Sahin, Zhang, and Arcuri]{sahin2026causes}
\BIBentryALTinterwordspacing
O.~Sahin, M.~Zhang, and A.~Arcuri, ``Causes and effects of fitness landscapes in system test generation: A replication study,'' \emph{Automated Software Engineering}, vol.~33, no.~8, 2026. [Online]. Available: \url{https://link.springer.com/article/10.1007/s10515-025-00539-z}
\BIBentrySTDinterwordspacing

\bibitem[Opsahl-Ong et~al.(2024)Opsahl-Ong, Ryan, Purtell, Broman, Potts, Zaharia, and Khattab]{opsahlong2024optimizinginstructionsdemonstrationsmultistage}
\BIBentryALTinterwordspacing
K.~Opsahl-Ong, M.~J. Ryan, J.~Purtell, D.~Broman, C.~Potts, M.~Zaharia, and O.~Khattab, ``Optimizing instructions and demonstrations for multi-stage language model programs,'' 2024. [Online]. Available: \url{https://arxiv.org/abs/2406.11695}
\BIBentrySTDinterwordspacing

\bibitem[Chen et~al.(2021)Chen, Tworek, Jun, Yuan, de~Oliveira~Pinto, Kaplan, Edwards, Burda, Joseph, Brockman, Ray, Puri, Krueger, Petrov, Khlaaf, Sastry, Mishkin, Chan, Gray, Ryder, Pavlov, Power, Kaiser, Bavarian, Winter, Tillet, Such, Cummings, Plappert, Chantzis, Barnes, Herbert-Voss, Guss, Nichol, Paino, Tezak, Tang, Babuschkin, Balaji, Jain, Saunders, Hesse, Carr, Leike, Achiam, Misra, Morikawa, Radford, Knight, Brundage, Murati, Mayer, Welinder, McGrew, Amodei, McCandlish, Sutskever, and Zaremba]{chen2021evaluating}
\BIBentryALTinterwordspacing
M.~Chen, J.~Tworek, H.~Jun, Q.~Yuan, H.~P. de~Oliveira~Pinto, J.~Kaplan, H.~Edwards, Y.~Burda, N.~Joseph, G.~Brockman, A.~Ray, R.~Puri, G.~Krueger, M.~Petrov, H.~Khlaaf, G.~Sastry, P.~Mishkin, B.~Chan, S.~Gray, N.~Ryder, M.~Pavlov, A.~Power, L.~Kaiser, M.~Bavarian, C.~Winter, P.~Tillet, F.~P. Such, D.~Cummings, M.~Plappert, F.~Chantzis, E.~Barnes, A.~Herbert-Voss, W.~Guss, A.~Nichol, A.~Paino, N.~Tezak, J.~Tang, I.~Babuschkin, S.~Balaji, S.~Jain, W.~Saunders, C.~Hesse, A.~N. Carr, J.~Leike, J.~Achiam, V.~Misra, E.~Morikawa, A.~Radford, M.~Knight, M.~Brundage, M.~Murati, K.~Mayer, P.~Welinder, B.~McGrew, D.~Amodei, S.~McCandlish, I.~Sutskever, and W.~Zaremba, ``Evaluating large language models trained on code,'' arXiv preprint arXiv:2107.03374, 2021. [Online]. Available: \url{https://arxiv.org/abs/2107.03374}
\BIBentrySTDinterwordspacing

\bibitem[Austin et~al.(2021)Austin, Odena, Nye, Bosma, Michalewski, Dohan, Jiang, Cai, Terry, Le, and Sutton]{austin2021program}
\BIBentryALTinterwordspacing
J.~Austin, A.~Odena, M.~Nye, M.~Bosma, H.~Michalewski, D.~Dohan, E.~Jiang, C.~Cai, M.~Terry, Q.~Le, and C.~Sutton, ``Program synthesis with large language models,'' arXiv preprint arXiv:2108.07732, 2021. [Online]. Available: \url{https://arxiv.org/abs/2108.07732}
\BIBentrySTDinterwordspacing

\bibitem[Hendrycks et~al.(2021)Hendrycks, Burns, Basart, Zou, Mazeika, Song, and Steinhardt]{hendrycks2021measuring}
\BIBentryALTinterwordspacing
D.~Hendrycks, C.~Burns, S.~Basart, A.~Zou, M.~Mazeika, D.~Song, and J.~Steinhardt, ``Measuring massive multitask language understanding,'' in \emph{Proceedings of the International Conference on Learning Representations (ICLR)}, 2021. [Online]. Available: \url{https://arxiv.org/abs/2009.03300}
\BIBentrySTDinterwordspacing

\bibitem[Deng et~al.(2024)Deng, Zhao, Tang, Gerstein, and Cohan]{deng2024investigating}
\BIBentryALTinterwordspacing
C.~Deng, Y.~Zhao, X.~Tang, M.~Gerstein, and A.~Cohan, ``Investigating data contamination in modern benchmarks for large language models,'' in \emph{Proceedings of the 2024 Conference of the North American Chapter of the Association for Computational Linguistics: Human Language Technologies (NAACL-HLT), Long Papers}, 2024, pp. 8706--8719. [Online]. Available: \url{https://aclanthology.org/2024.naacl-long.482/}
\BIBentrySTDinterwordspacing

\bibitem[Sahu et~al.(2022)Sahu, Rodriguez, Laradji, Atighehchian, Vazquez, and Bahdanau]{sahu2022dataintentclassification}
\BIBentryALTinterwordspacing
G.~Sahu, P.~Rodriguez, I.~Laradji, P.~Atighehchian, D.~Vazquez, and D.~Bahdanau, ``Data augmentation for intent classification with off-the-shelf large language models,'' in \emph{Proceedings of the 4th Workshop on NLP for Conversational AI}, B.~Liu, A.~Papangelis, S.~Ultes, A.~Rastogi, Y.-N. Chen, G.~Spithourakis, E.~Nouri, and W.~Shi, Eds.\hskip 1em plus 0.5em minus 0.4em\relax Dublin, Ireland: Association for Computational Linguistics, 2022, pp. 47--57. [Online]. Available: \url{https://aclanthology.org/2022.nlp4convai-1.5}
\BIBentrySTDinterwordspacing

\bibitem[Jiang et~al.(2023)Jiang, Wu, Lin, Yang, and Qiu]{jiang2023llmlingua}
H.~Jiang, Q.~Wu, C.-Y. Lin, Y.~Yang, and L.~Qiu, ``Llmlingua: Compressing prompts for accelerated inference of large language models,'' in \emph{Proceedings of the 2023 Conference on Empirical Methods in Natural Language Processing (EMNLP)}.\hskip 1em plus 0.5em minus 0.4em\relax Singapore: Association for Computational Linguistics, Dec. 2023, pp. 13\,358--13\,376.

\bibitem[Kwon et~al.(2023)Kwon, Li, Zhuang, Sheng, Zheng, Yu, Gonzalez, Zhang, and Stoica]{kwon2023efficient}
W.~Kwon, Z.~Li, S.~Zhuang, Y.~Sheng, L.~Zheng, C.~H. Yu, J.~E. Gonzalez, H.~Zhang, and I.~Stoica, ``Efficient memory management for large language model serving with pagedattention,'' in \emph{Proceedings of the 29th Symposium on Operating Systems Principles (SOSP '23)}, 2023, pp. 611--626.

\bibitem[Leviathan et~al.(2023)Leviathan, Kalman, and Matias]{leviathan2023fast}
Y.~Leviathan, M.~Kalman, and Y.~Matias, ``Fast inference from transformers via speculative decoding,'' in \emph{Proceedings of the 40th International Conference on Machine Learning}, vol. 202, 2023, pp. 19\,274--19\,286.

\bibitem[Bang(2023)]{bang2023gptcache}
\BIBentryALTinterwordspacing
F.~Bang, ``{GPTC}ache: An open-source semantic cache for {LLM} applications enabling faster answers and cost savings,'' in \emph{Proceedings of the 3rd Workshop for Natural Language Processing Open Source Software (NLP-OSS 2023)}, L.~Tan, D.~Milajevs, G.~Chauhan, J.~Gwinnup, and E.~Rippeth, Eds.\hskip 1em plus 0.5em minus 0.4em\relax Singapore: Association for Computational Linguistics, Dec. 2023, pp. 212--218. [Online]. Available: \url{https://aclanthology.org/2023.nlposs-1.24}
\BIBentrySTDinterwordspacing

\bibitem[Baghdadi et~al.(2021)Baghdadi, Merouani, Leghettas, Abdous, Arbaoui, BENATCHBA, and amarasinghe]{baghdadi2021dlcostmodel}
\BIBentryALTinterwordspacing
R.~Baghdadi, M.~Merouani, M.-H. Leghettas, K.~Abdous, T.~Arbaoui, K.~BENATCHBA, and S.~amarasinghe, ``A deep learning based cost model for automatic code optimization,'' in \emph{Proceedings of Machine Learning and Systems}, A.~Smola, A.~Dimakis, and I.~Stoica, Eds., vol.~3, 2021, pp. 181--193. [Online]. Available: \url{https://proceedings.mlsys.org/paper_files/paper/2021/file/d9387b6d643efb25132be36f7b908d96-Paper.pdf}
\BIBentrySTDinterwordspacing

\bibitem[Lamb and Zacchiroli(2022{\natexlab{a}})]{lamb2022reproducible}
\BIBentryALTinterwordspacing
C.~Lamb and S.~Zacchiroli, ``Reproducible builds: Increasing the integrity of software supply chains,'' \emph{IEEE Software}, vol.~39, no.~2, pp. 62--70, 2022. [Online]. Available: \url{https://doi.org/10.1109/MS.2021.3073045}
\BIBentrySTDinterwordspacing

\bibitem[de~Carn{\'e}~de Carnavalet and Mannan(2014)]{decarnedecarnelatmannan2014challenges}
\BIBentryALTinterwordspacing
X.~de~Carn{\'e}~de Carnavalet and M.~Mannan, ``Challenges and implications of verifiable builds for security-critical open-source software,'' in \emph{Proceedings of the 30th Annual Computer Security Applications Conference (ACSAC ’14)}, 2014, pp. 16--25. [Online]. Available: \url{https://dl.acm.org/doi/10.1145/2664243.2664288}
\BIBentrySTDinterwordspacing

\bibitem[de~Carn\'{e}~de Carnavalet and Mannan(2014)]{carnavalet2014verifiablebuilds}
\BIBentryALTinterwordspacing
X.~de~Carn\'{e}~de Carnavalet and M.~Mannan, ``Challenges and implications of verifiable builds for security-critical open-source software,'' in \emph{Proceedings of the 30th Annual Computer Security Applications Conference}, ser. ACSAC '14.\hskip 1em plus 0.5em minus 0.4em\relax New York, NY, USA: Association for Computing Machinery, 2014, p. 16–25. [Online]. Available: \url{https://doi.org/10.1145/2664243.2664288}
\BIBentrySTDinterwordspacing

\bibitem[Renze and Guven(2024)]{renze2024effectsamplingtemperatureproblem}
\BIBentryALTinterwordspacing
M.~Renze and E.~Guven, ``The effect of sampling temperature on problem solving in large language models,'' 2024. [Online]. Available: \url{https://arxiv.org/abs/2402.05201}
\BIBentrySTDinterwordspacing

\bibitem[Shi et~al.(2022)Shi, Wen, Cogo, Chen, and Jiang]{shi2022verifiablebuildslargescalecomm}
Y.~Shi, M.~Wen, F.~R. Cogo, B.~Chen, and Z.~M. Jiang, ``An experience report on producing verifiable builds for large-scale commercial systems,'' \emph{IEEE Transactions on Software Engineering}, vol.~48, no.~9, pp. 3361--3377, 2022.

\bibitem[Lamb and Zacchiroli(2022{\natexlab{b}})]{lamb2022reproduciblebuildssupplechain}
C.~Lamb and S.~Zacchiroli, ``Reproducible builds: Increasing the integrity of software supply chains,'' \emph{IEEE Software}, vol.~39, no.~2, pp. 62--70, 2022.

\bibitem[Chen et~al.(2022)Chen, Wen, Shi, Lin, Rajbahadur, and Jiang]{chen2022reproducibledl}
\BIBentryALTinterwordspacing
B.~Chen, M.~Wen, Y.~Shi, D.~Lin, G.~K. Rajbahadur, and Z.~M.~J. Jiang, ``Towards training reproducible deep learning models,'' in \emph{Proceedings of the 44th International Conference on Software Engineering}, ser. ICSE '22.\hskip 1em plus 0.5em minus 0.4em\relax New York, NY, USA: Association for Computing Machinery, 2022, p. 2202–2214. [Online]. Available: \url{https://doi.org/10.1145/3510003.3510163}
\BIBentrySTDinterwordspacing

\bibitem[Chisnall(2013)]{chisnall2013challenge}
D.~Chisnall, ``The challenge of cross-language interoperability,'' \emph{Communications of the ACM}, vol.~56, no.~12, pp. 50--56, 2013.

\bibitem[Grimmer et~al.(2018)Grimmer, Schatz, Seaton, Würthinger, and Luján]{grimmer2018crosslanguage}
M.~Grimmer, R.~Schatz, C.~Seaton, T.~Würthinger, and M.~Luján, ``Cross-language interoperability in a multi-language runtime,'' \emph{ACM Transactions on Programming Languages and Systems}, vol.~40, no.~2, p.~8, 2018.

\bibitem[Liu et~al.(2020)Liu, Chen, Zhang, Qin, Ji, Lin, and Yang]{liu2020enhancinginteropdl}
Y.~Liu, C.~Chen, R.~Zhang, T.~Qin, X.~Ji, H.~Lin, and M.~Yang, ``Enhancing the interoperability between deep learning frameworks by model conversion,'' in \emph{Proceedings of the 28th ACM Joint Meeting on European Software Engineering Conference and Symposium on the Foundations of Software Engineering}, ser. ESEC/FSE 2020.\hskip 1em plus 0.5em minus 0.4em\relax New York, NY, USA: Association for Computing Machinery, 2020, p. 1320–1330.

\bibitem[Jin et~al.(2020)Jin, Bercea, Le, Chen, Su, Imai, Negishi, Leu, O'Brien, Kawachiya, and Eichenberger]{jin2020compilingonnxneuralnetwork}
\BIBentryALTinterwordspacing
T.~Jin, G.-T. Bercea, T.~D. Le, T.~Chen, G.~Su, H.~Imai, Y.~Negishi, A.~Leu, K.~O'Brien, K.~Kawachiya, and A.~E. Eichenberger, ``Compiling onnx neural network models using mlir,'' 2020. [Online]. Available: \url{https://arxiv.org/abs/2008.08272}
\BIBentrySTDinterwordspacing

\bibitem[Brambilla et~al.(2017)Brambilla, Cabot, and Wimmer]{brambilla2017model}
M.~Brambilla, J.~Cabot, and M.~Wimmer, \emph{Model-Driven Software Engineering in Practice}, 2nd~ed., ser. Synthesis Lectures on Software Engineering.\hskip 1em plus 0.5em minus 0.4em\relax Morgan \& Claypool Publishers, 2017.

\bibitem[Garlan et~al.(2009)Garlan, Cheng, and Schmerl]{garlan2009software}
D.~Garlan, S.-W. Cheng, and B.~Schmerl, ``Software architecture-based self-adaptation,'' in \emph{Autonomic Computing and Networking}.\hskip 1em plus 0.5em minus 0.4em\relax Springer, 2009, pp. 31--55.

\bibitem[de~Lemos et~al.(2013)de~Lemos, Giese, M{\"u}ller, Shaw, Andersson, Litoiu, Schmerl, Tamura, Villegas, Vogel, Weyns, Baresi, Becker, Bencomo, Brun, Cukic, Desmarais, Dustdar, Engels, Geihs, G{\"o}schka, Gorla, Grassi, Inverardi, Karsai, Kramer, Lopes, Magee, Malek, Mankovskii, Mirandola, Mylopoulos, Nierstrasz, Pezz{\`e}, Prehofer, Sch{\"a}fer, Schlichting, Smith, Sousa, Tahvildari, Wong, and Wuttke]{delemos2013software}
R.~de~Lemos, H.~Giese, H.~A. M{\"u}ller, M.~Shaw, J.~Andersson, M.~Litoiu, B.~Schmerl, G.~Tamura, N.~M. Villegas, T.~Vogel, D.~Weyns, L.~Baresi, B.~Becker, N.~Bencomo, Y.~Brun, B.~Cukic, R.~Desmarais, S.~Dustdar, G.~Engels, K.~Geihs, K.~M. G{\"o}schka, A.~Gorla, V.~Grassi, P.~Inverardi, G.~Karsai, J.~Kramer, A.~Lopes, J.~Magee, S.~Malek, S.~Mankovskii, R.~Mirandola, J.~Mylopoulos, O.~Nierstrasz, M.~Pezz{\`e}, C.~Prehofer, W.~Sch{\"a}fer, R.~Schlichting, D.~B. Smith, J.~P. Sousa, L.~Tahvildari, K.~Wong, and J.~Wuttke, ``Software engineering for self-adaptive systems: A second research roadmap,'' in \emph{Software Engineering for Self-Adaptive Systems II}, ser. Lecture Notes in Computer Science.\hskip 1em plus 0.5em minus 0.4em\relax Springer, 2013, vol. 7475, pp. 1--32.

\bibitem[Kephart and Chess(2003)]{kephart2003vision}
J.~O. Kephart and D.~M. Chess, ``The vision of autonomic computing,'' \emph{Computer}, vol.~36, no.~1, pp. 41--50, 2003.

\bibitem[Sanwouo~Chekam et~al.(2025)Sanwouo~Chekam, Quinton, and Temple]{chekam2025breaking}
\BIBentryALTinterwordspacing
B.~P. Sanwouo~Chekam, C.~Quinton, and P.~Temple, \emph{Breaking the Loop: AWARE is the New MAPE-K}.\hskip 1em plus 0.5em minus 0.4em\relax New York, NY, USA: Association for Computing Machinery, 2025, p. 626–630. [Online]. Available: \url{https://doi.org/10.1145/3696630.3728512}
\BIBentrySTDinterwordspacing

\bibitem[Benavides et~al.(2010)Benavides, Segura, and Ruiz-Cort{\'e}s]{benavides2010automated}
D.~Benavides, S.~Segura, and A.~Ruiz-Cort{\'e}s, ``Automated analysis of feature models 20 years later: A literature review,'' \emph{Information Systems}, vol.~35, no.~6, pp. 615--636, 2010.

\bibitem[Th{\"u}m et~al.(2014)Th{\"u}m, K{\"a}stner, Benduhn, Meinicke, Saake, and Leich]{thum2014featureide}
T.~Th{\"u}m, C.~K{\"a}stner, F.~Benduhn, J.~Meinicke, G.~Saake, and T.~Leich, ``{FeatureIDE}: An extensible framework for feature-oriented software development,'' \emph{Science of Computer Programming}, vol.~79, pp. 70--85, 2014.

\end{thebibliography}

\end{document}